\documentclass[journal=jpccck,manuscript=article]{achemso}
\usepackage[version=3]{mhchem} 
\usepackage{xr}
\usepackage{url}
\usepackage{graphicx}
\usepackage{subfigure}
\usepackage{epstopdf}
\usepackage{dcolumn}
\usepackage{bm}
\usepackage{color}
\usepackage{float}
\usepackage{scalerel}
\usepackage[caption = false]{subfig}
\usepackage{amsfonts} 
\usepackage[toc, page] {appendix}
\SectionNumbersOn

\makeatletter

\title{Dyadic Green Function Approach to Multichromophoric Forster Resonance Energy Transfer under Electromagnetic Fluctuations near Metallic Thin Films}%

\author{Changhao Meng}
\altaffiliation{Contributed equally to this work}
\affiliation[XJTU-phys]
{Department of Physics, State Key Laboratory of Surface Physics, Institute of Nanoelectronic Devices and Quantum Computing and Key Laboratory of Micro and Nano Photonic Structures (Ministry of Education), Fudan University, Shanghai 200433, China}
\author{Xin Chen}
\altaffiliation{Contributed equally to this work}
\email{xin.chen.nj@xjtu.edu.cn}
\affiliation[XJTU]
{Center of Nanomaterials for Renewable Energy, State Key Laboratory of Electrical Insulation and Power Equipment, School of Electrical Engineering, Xi'an Jiaotong University, Xi'an 710054, Shaanxi, China}

\author{Zhenghua An}
\affiliation[fudan]{State Key Laboratory of Surface Physics, Department of Physics and Collaborative Innovation Center of Advanced Microstructures, Fudan University, Shanghai 200433, PR China}

\begin{document}

\maketitle

\begin{abstract}
The near-field spectroscopic information is critically important to determine the Forster resonant energy transfer(FRET) rate and the distance dependence in the vicinity of metal surfaces. The high density of evanescent near-field modes in the vicinity of a metal surface can strongly modulate the FRET in multichromophoric systems. Based on the previous generalized FRET [A. Poudel, X. Chen and M. Ratner, J. Phys. Chem. Lett. 7(2016) 955], the theory of FRET is generalized for the multichromophore aggregates and nonequilibrium situations in the vicinity of evanescent surface electromagnetic waves of nanophotonic structures. The classic dyadic green function (DGF) approach to multichromophoric
FRET (MC-FRET) in the existence of evanescent near-field is established. The classic DGF approach provides a microscopic understanding of the interaction between the emission and absorption spectral coupling and the evanescent electromagnetic field. 
The MC-FRET of the ring structures demonstrates complicated distance dependence in the vicinity of the silver thin film. Given the analytic expression of hyperbolic multi-layer thin film, the generalized coupling due to the metallic thin film is determined by the scattering DGF above the metal surface. The decomposition of the reduced scattering DGF by ignoring $S$ wave in the evanescent wave shows how the interface of the metallic thin films modulates the MC-FRET. 
\end{abstract}

\maketitle

\section{Introduction}
Excitonic energy transfer (EET) exists in many fields of physics, chemistry, and biology and has received a lot of attention\cite{intro1, intro3, intro4}. It is a fundamental problem in various physical and chemical processes. In general, the EET rate can be well described by the Förster resonance energy transfer (FRET) theory. The enhancement of FRET efficiency \cite{meng2019forster,PCRET-hole} can lead to an increase in FRET distance which has been applied in the biological macromolecule spectroscopy research, fluorescence imaging, biosensing, macromolecular conformation research, DNA detection\cite{PEF,PCRET-application}, {\it{etc}}. In the FRET system with the donor and acceptor, the coupling between them is much weaker than the system-bath coupling. Therefore, both the acceptor and donor can be treated as electrostatic point dipoles. EET is critically important to solar energy harvesting and energy conversion. In solar power electronics and photonics materials,  the interaction between the exciton and the evanescent near-field electromagnetic waves of the electrodes demands further studies. Understanding how to optimize the electrodes of solar cells\cite{solar,solar2} can provide a new approach to improve the efficiencies of energy conversion and transport. 

Aspects of modern research on EET in photosynthetic light-harvesting systems have focused on energy transfer as a coherent collective phenomenon\cite{photo1, photo2, photo3}. This feature has been highlighted as central to several transfer mechanisms, such as super transfer and a network renormalization scheme, and predicts dramatic enhancements of energy transfer rates. Qualitative arguments explaining such behavior often rely on interactions between donors and acceptors that induce excitation delocalization and establish quantum correlations, such as entanglement, between chromophores. Consequently, this observed unexpected rate enhancement has been widely attributed to the quantum coherence of acceptors and donors. FRET is a photophysical process where the electronic excitation energy is transferred from an excited donor chromophore to the near acceptor chromophores by the dipole-dipole non-radiative interactions. Multichromophore(MC) systems are building blocks of molecular optoelectronic devices\cite{MC-struc}. The conventional FRET theory only considers the non-radiative energy transfer of a sole donor-acceptor pair. Unfortunately, in a molecular chromophore aggregator,  a single acceptor rarely exists. The FRET theory significantly underestimates the energy transfer rate in the MC systems. Therefore, the multichromophoric FRET (MC-FRET) theory was developed to solve this problem\cite{silbey1}.

In the conventional MC-FRET, the system-bath coupling makes the resonant energy transfer incoherent\cite{noise2}. At the same time, the fluctuating currents in the metal give rise to noisy electromagnetic fields in the surrounding region\cite{noise}. The electromagnetic field in the space around bodies is stochastic due to quantum and thermal fluctuations. The evanescent near-filed is also an incoherent thermal bath. The MC-FRET are derived classically according to the electrostatic dipole-dipole interaction\cite{brumer} at the near field approximation in a vacuum. The current MC-FRET doesn't include the effect of evanescent EM waves in the existence of metal/nanophotonic structures. The current MC-FRET theory\cite{silbey1,brumer} doesn't consider the presence of the evanescent near-field at the nanoscale. The interaction of the MC structure and the nonequilibrium thin-film evanescent EM field demands the extension of the current MC-FRET. In the metal/nanophotonic structures, the coupling of evanescent near-field and the donor and acceptor MC structures are accounted for in DGF. As an extension of the orientation factor in the conventional FRET, the coupling factor based on DGF\cite{CX,meng2019forster,george} is used for the extension of MC-FRET in the vicinity of metals. The evanescent near-field EM field is characterized by the cross-spectral density tensor\cite{scoh}. The spatial coherence of the evanescent field can last for several wavelengths. The nanophotonic metasurface can modulate the evanescent near field\cite{NFModulation}, {\it{i.e.}} it can change the corresponding local density of states (LDOS) and cross-spectral density tensor. In the existence of nanophotonic materials, the LDOS modulation can strongly affect the emission intensity and radiative decay time in the excitation energy dynamics. The evanescent near-field EM waves can enhance the conductivity of organic semiconductors \cite{Hybrid-conductivity}, optical chemical reaction\cite{Hybrid-Chemistry}, and {\it{etc}.}  The hybridization of the exciton and evanescent vacuum field to promote the non-radiative energy transfer\cite{CX,meng2019forster,Hybrid-Energy}. In the existence of the evanescent near-field, the FRET rate in the chromophore aggregates\cite{CX,meng2019forster} can be tuned. 
The effect of evanescent near-fields is included in the dyadic Green functions{DGF}. As an extension of the orientation factor in the conventional FRET, the coupling factor is based on DGF\cite{CX,meng2019forster,george}. How to evaluate DGF numerically is essential to study the generalized FRET. To understand how to tune and improve the FRET efficiency, we need to know how to modulate the evanescent near-field with nanophotonic structures to match the emission and absorption spectral overlap in FRET\cite{meng2019forster}. Within the multiple acceptors, it is assumed that the resonant energy is either released from the donor or transferred to only one of the acceptors. Multichromophore(MC) systems are building blocks of molecular optoelectronic devices\cite{MC-struc}. In the FRET system with the donor and acceptor, the coupling between them is much weaker than the system-bath coupling. Both the acceptor and donor can be treated as electrostatic point dipoles. The FRET theory significantly underestimates the energy transfer rate in the MC systems. Therefore, the multichromophoric F{\"o}rster resonance energy transfer (MC-FRET) theory was developed to solve this problem\cite{silbey1}.

In this paper, we present the generalized MC-FRET approach in vicinity of thin metallic film. The paper is organized into four sections. 
In Section~\ref{method}, we discuss the generalized MC-FRET formula in the existence of evanescent EV waves. In the near-field region, the generalized MC-FRET can be reduced to the conventional MC-FRET. 
In Section~\ref{MCS}, we discuss the distance dependence of two MC structures, the one-to-two and three-fold ring-ring structures. 
In Section~\ref{mod}, we discuss how the scattering DGF module MC-FRET and the polarization orientation dependence. 
In Section~\ref{cr},  we give concluding remarks and conclude this paper by discussing the implications of our results and the future work.

\section{Generalized Multichromophoric FRET under Evanescent Electromagnetic Field}\label{method}
The Forster resonant energy transfer (FRET) in the vicinity of the metal and nanophotonics surface is established\cite{CX} previously based on the dyadic Green function. From the classical electromagnetic theory, the energy flux density of the electromagnetic field from the donor to acceptor is given by the Poynting vector, $\langle \mathbf{S} \rangle_p=\langle \mathbf{E} \times \mathbf{H}\rangle_p$. Adopting the classical perspective by Silbey and co-workers\cite{silbey2,silbey3}, the energy transfer from the donor to acceptor can be described as the two coupled oscillating dipoles. 
FRET under the influence of the thermal nonequilibrium evanescent near field as electromagnetic fluctuations can be derived based on the dyadic Green function (DGF) and coupling factor\cite{CX}. The energy flow $Q(t)$ in the time domain is define as,
\begin{equation}
    \dot{Q}(t)= \mathbf{E}(\mathbf{r}_{D}, \mathbf{r}_{A},t)\cdot \dot{\mathbf{p}}_{A}(t),
\end{equation}
With the Fourier transformation, the FRET rate can be reformed to be in the frequency domain,
\begin{eqnarray}\label{rate}
    \tilde{\dot{Q}} (0)& = &  -i \int_{-\infty}^{\infty} d\omega \omega  \mathbf{E}(\mathbf{r}_{A}, \mathbf{r}_{D}, -\omega)\cdot {\mathbf{p}}_{A}(\omega) \\ \nonumber 
    &=&\int^{\infty}_{-\infty} d\omega\,  \omega^4 \mu_0^2 \, \sigma_A(\omega)\, \sigma_D(\omega)\, \cdot  |\mathbf{n}_A \cdot \hat{G}(\mathbf{r}_A, \mathbf{r}_D, \omega) \cdot \mathbf{n}_D|^2\,,
\end{eqnarray}
where the absorption spectrum of the acceptor chromophore  $\sigma_A (\omega)$ and the emission spectrum of the donor chromophore $\sigma_D(\omega)$ and the unit dipoles of donor and acceptor, $\mathbf{n}_D$ and $\mathbf{n}_A$, and
the retarded photon Green's function $\hat{G}(\mathbf{r}_A, \mathbf{r}_D, \omega)$ satisfies the following wave equation,
\begin{align}
     & \Big[\partial_{i}\partial_{j} -\delta_{ij}\Big( \nabla^{2}+\frac{\omega^{2}\varepsilon(
            \mathbf{r}, \omega)  }{c^{2}}\Big)\Big]
    G_{ik}(\mathbf{r}_D,\mathbf{r}_A, \omega)  \nonumber                                             \\
     & = \delta^{3}\left(\mathbf{r}_D-\mathbf{r}_A\right)  \delta_{jk}\,,
\end{align}
where $c$ is the speed of light in vacuum and $\mathbf{r}_D$ and $\mathbf{r}_A$ are the donor and acceptor positions, respectively. The details of the derivation can be found in our previous work\cite{CX}.  
In FRET, the effect of electromagnetic environment is fully captured by DGF, which can be obtained computationally by solving Maxwell's equations.
DGF can be used to describe the electric field $\mathbf{E}_D(\mathbf{r},\mathbf{r}_D, \omega)$ of the donor dipole $\mathbf{p}_{D} (\omega)$, where $p_D(\omega)$ is the strength or magnitude of the electric dipole and $\mathbf{n}_D$ is the unit dipole, located at $\mathbf{r}_D$ in the presence of arbitrary metallic environment,
\begin{align}\label{DGF2EM}
    \mathbf{E}_D(\mathbf{r},\mathbf{r}_D,  \omega) = \omega^2  \mu_0 \hat{\mathbf{G}} (\mathbf{r}, \mathbf{r}_D,  \omega) \cdot \mathbf{p}_{D}(\omega) = \omega^2  \mu_0 p_D(\omega) \hat{\mathbf{G}} (\mathbf{r}, \mathbf{r}_D, \omega) \cdot \mathbf{n}_D\,
\end{align}
where $ \mu_0$ is the vacuum permeability.

For the MC structures in the existence of the evanescent near field, the energy flow in time domain $Q(t)$ can be generalized as,
\begin{equation}
    \dot{Q}(t)= \sum_{n=1}^{N_D} \sum_{m=1}^{N_A} \mathbf{E}(\mathbf{r}_{A_m}, \mathbf{r}_{D_n},t)\cdot \dot{\mathbf{p}}_{A_m}(t),
\end{equation}
where $N_D$ is the number of MC donors, $N_A$ is the number of MC acceptors, $\mathbf{E}( \mathbf{r}_{A_m},\mathbf{r}_{D_n},t)$ is the electric field vector at $\mathbf{r}_{A_m}$ of the m-th acceptor chromophore due to the n-th donor chromophore at $\mathbf{r}_{D_n}$, and $\mathbf{p}_{A_m}(t)$ is the polarizability vector of the m-th acceptor chromophore. After transforming $Q(t)$ into the frequency domain, the MC-FRET rate is defined as, 
\begin{equation}
  \tilde{\dot{Q}} (0)= -i \int_{-\infty}^{\infty} d\omega \omega \sum_{n=1}^{N_D} \sum_{m=1}^{N_A} \mathbf{E}^*(\mathbf{r}_{A_m}, \mathbf{r}_{D_n}, -\omega)\cdot {\mathbf{p}}_{A_m}(\omega)
\end{equation}
where ${\mathbf{p}}_{A_m}(\omega)$ is the polarizability of the m-th acceptor chromophore molecule in the frequency domain and $\sum_{n=1}^{N_D} \mathbb{E}(\mathbf{r}_{A_m},\mathbf{r}_{D_n}, \omega)$ is the total electric field at the position of the m-th acceptor due to the $N_D$ chromophore donors. The electric field, $\mathbb{E}(\mathbf{r}_{A_m}, \omega )$  at the m-th acceptor can be formulated to be,
\begin{eqnarray}\label{E}
    \mathbb{E}(\mathbf{r}_{A_m}, \omega )
    =  \mu_0  \omega^2  \sum_{m=1}^{N_A}  \left ( \sum_{k=1,k\neq m}^{N_A} \hat{\mathbf{G}} (\mathbf{r}_{A_m}, \mathbf{r}_{A_k}, \omega) \cdot \mathbf{p}_{{A_k}}(\omega)
    +   \sum_{l=1}^{N_D} \hat{\mathbf{G}}(\mathbf{r}_{A_m},\mathbf{r}_{D_l}, \omega) \cdot \mathbf{p}_{D_l}(\omega) \right ).
\end{eqnarray}
The generalized MC-FRET rate is defined in the frequency domain as,
\begin{eqnarray}\label{Q}
    \tilde{\dot{Q}} (0)
    & = & -i \mu_{0} \int_{-\infty}^{\infty}\; d\omega\; \omega^3 \sum_{m=1}^{N_A} \Bigg( \sum_{k=1,k\neq m}^{N_A}
    \hat{\mathbf{G}}^{*} (\mathbf{r}_{A_m}, \mathbf{r}_{A_k},\omega) \cdot  \mathbf{p}_{{A_k}}^{*}(\omega) \\ \nonumber
    & + &   \sum_{l=1}^{N_D}
     \hat{\mathbf{G}}^{*} (\mathbf{r}_{A_m},\mathbf{r}_{D_l}, \omega) \cdot \mathbf{p}_{D_l}^{*}(\omega) \Bigg) \cdot {\mathbf{p}}_{A_m}(\omega) .
\end{eqnarray}
In the vector compact form, the MC-FRET rate can be expressed as, 
\begin{eqnarray}
    \tilde{\dot{Q}} (0)
    =  -i \mu_{0}  \int_{-\infty}^{\infty} d\omega \omega^3 \left (   \hat{\mathbf{G}}^{\mathrm{A}*} (\omega) \cdot \mathbf{P}_{\mathrm{A}}^*(\omega)  +  \hat{\mathbf{G}}^{\mathrm{AD}*} (\omega) \cdot \mathbf{P}_{\mathrm{D}}^*(\omega) \right ) \cdot \mathbf{P}_{\mathrm{A}}(\omega) 
\end{eqnarray}
where $\mathbf{P}_{\mathrm{A}}(\omega)=\{\mathbf{p}_{A_1}^T(\omega), \mathbf{p}_{A_2}^T(\omega),\cdots,\mathbf{p}_{A_N}^T(\omega) \}^T$ and $\mathbf{P}_{\mathrm{D}}(\omega)=\{\mathbf{p}_{D_1}^T(\omega), \mathbf{p}_{D_2}^T(\omega),\cdots,\mathbf{p}_{D_M}^T(\omega) \}^T$ are the column vectors for the MC donor and acceptor aggregates, $\hat{\mathbf{G}}^{\mathrm{A}}(\omega)$ is the matrix of dyadic Green functions, whose element in the $i^{th}$ row and $j^{th}$ column ($3 \times 3$ submatrix) is $\hat{\mathbf{G}}^{\mathrm{A}}_{ij}(\omega)=\hat{\mathbf{G}} (\mathbf{r}_{A_i}, \mathbf{r}_{A_j},\omega)$, $\hat{\mathbf{G}}^{\mathrm{AD}}(\omega)$ is also the matrix of dyadic Green function, whose element in the $i^{th}$ row and $j^{th}$ column ($3 \times 3$ submatrix) is  $\hat{\mathbf{G}}^{\mathrm{AD}}_{ij}(\omega) = \hat{\mathbf{G}} (\mathbf{r}_{A_i}, \mathbf{r}_{D_j},\omega)$. For $\hat{\mathbf{G}}^{\mathrm{A}}(\omega)$, its diagonal dyadic Green functions, $\hat{\mathbf{G}}^{\mathrm{A}}_{ii}(\omega)$ are zero matrices. DGF and induced polarizability have the conjugate symmetry properties, $\mathbf{p}_{A_i}(-\omega)=\mathbf{p}_{A_i}^*(\omega)$, $\mathbf{p}_{D_j}(-\omega)=\mathbf{p}_{D_j}^*(\omega)$, $\hat{\mathbf{G}}^{\mathrm{A}}(-\omega) = \hat{\mathbf{G}}^{\mathrm{A}*}(\omega) $, $\hat{\mathbf{G}}^{\mathrm{D}}(-\omega) = \hat{\mathbf{G}}^{\mathrm{D}*}(\omega) $, $\hat{\mathbf{G}}^{\mathrm{DA}}(-\omega) = \hat{\mathbf{G}}^{\mathrm{DA}*}(\omega) $, $\hat{\mathbf{G}}^{\mathrm{AD}}(-\omega) = \hat{\mathbf{G}}^{\mathrm{AD}*}(\omega) $, and ${\boldsymbol{\chi}}_{\mathrm{A}}(-\omega)= {\boldsymbol{\chi}}_{\mathrm{A}}^*(\omega)$. Therefore, 
The induced polarizability of the m-th acceptor chromophore in the MC structure can be expressed as,
\begin{eqnarray}
    \mathbf{p}_{\mathrm{A}_m}(\omega) & = & \epsilon_{0}  \boldsymbol{{\chi}}_{\mathrm{A}_{m}}(\omega)\omega^{2} \mu_{0} \Bigg( \sum_{k = 1,k\neq m}^{N_{A}} \hat{\mathbf{G}} (\mathbf{r}_{\mathrm{A}_m},\mathbf{r}_{\mathrm{A}_k}, \omega)\cdot \mathbf{p}_{\mathrm{A}_{k}}(\omega)   \\ \nonumber
    && + \sum_{n=1}^{N_{\mathrm{D}}} \hat{\mathbf{G}}\left(\mathbf{r}_{\mathrm{A}_m},\mathbf{r}_{\mathrm{D}_l},\omega\right) \cdot\mathbf{p}_{\mathrm{D}_l}(\omega) \Bigg),
\end{eqnarray}
where $\boldsymbol{\chi}_{\mathrm{A}_{m}}(\omega) $ is the polarizability tensor of the i-th acceptor chromophore, $\hat{\mathbf{G}} (\mathbf{r}_{A_m}, \mathbf{r}_{A_k},\omega)$ is the DGF between the m-th and k-th acceptor chromophores, and $\hat{\mathbf{G}}(\mathbf{r}_{A_m},\mathbf{r}_{D_l}, \omega)$ is the DGF between the l-th donor and m-th acceptor chromophores. 
The induced polarizability of the acceptor aggregate, $\mathbf{P}_A(\omega)$ in the frequency domain can be expressed in compact vector form as,
\begin{eqnarray}\label{PDA}
    \mathbf{P}_A(\omega) &  =  & \epsilon_{0}\omega^{2} {\boldsymbol{\chi}}_{\mathrm{A}}(\omega) \mu_{0}
    \left ( \hat{\mathbf{G}}^{\mathrm{A}}(\omega) \cdot \mathbf{P}_{\mathrm{A}}(\omega)   + \hat{\mathbf{G}}^{\mathrm{AD}}(\omega) \cdot \mathbf{P}_{\mathrm{D}}(\omega) \right )\\ \nonumber
    & = & \mathbf{K}(\omega)^{-1} \cdot \hat{\mathbf{G}}^{\mathrm{AD}}(\omega) \cdot \mathbf{P}_{\mathrm{D}}(\omega),
\end{eqnarray}
where $\mathbf{K}(\omega) = (\frac{c^2} {\omega^2 \boldsymbol{\chi}_A }-\hat{\mathbf{G}}^{\mathrm{A}}(\omega))$, and 
\begin{equation}
    {\boldsymbol{\chi}}_{\mathrm{A}} (\omega) =
    \begin{bmatrix}
        {\boldsymbol{\chi}}_{\mathrm{A}_{1}}(\omega)   & 0   & \dots  & 0   \\
        0   & {\boldsymbol{\chi}}_{\mathrm{A}_{2}}(\omega)   & \dots  & 0   \\
        \vdots                                         & \vdots                                         & \vdots & \vdots                                         \\
        0 & 0 & \dots  & {\boldsymbol{\chi}}_{\mathrm{A}_{N_A}}(\omega) \\
    \end{bmatrix}.
\end{equation}
As a result, the generalized MC-FRET rate can be expressed as,
\begin{eqnarray}\label{Qvector}
    \tilde{\dot{Q}} (0)
    &=& -i \mu_{0}  \int_{-\infty}^{\infty}\; d\omega\; \omega^3 \; \hat{\mathbf{G}}^{\mathrm{A}*} (\omega) \cdot \mathbf{P}_{\mathrm{A}}^*(\omega)  \cdot \mathbf{P}_{\mathrm{A}}(\omega) +  \\ \nonumber
    && -i \mu_{0} \int_{-\infty}^{\infty} \; d\omega \; \omega^3 \;\hat{\mathbf{G}}^{\mathrm{AD}*} (\omega) \cdot \mathbf{P}_{\mathrm{D}}^*(\omega) \cdot \mathbf{K}(\omega)^{-1} \cdot \hat{\mathbf{G}}^{\mathrm{AD}}(\omega) \cdot \mathbf{P}_{\mathrm{D}}(\omega),
\end{eqnarray}
In reference to the conventional MC-FRET rate\cite{CX}, the generalized MC-FRET rate is defined as, 
\begin{eqnarray}\label{Qvectorfinal}
    \gamma_{MC}
    &= & 2 \mu_0 \int_{0}^{\infty} d\omega \;  \omega^3  \; \textbf{Im}  \Big(  \hat{\mathbf{G}}^{\mathrm{A}*} (\omega) \cdot \mathbf{P}_{\mathrm{A}}^*(\omega) \cdot \mathbf{P}_{\mathrm{A}}(\omega) \\ \nonumber
    & + & \hat{\mathbf{G}}^{\mathrm{AD}*} (\omega) \cdot \mathbf{P}_{\mathrm{D}}^*(\omega)\cdot \mathbf{K}(\omega)^{-1} \cdot \hat{\mathbf{G}}^{\mathrm{AD}}(\omega) \cdot \mathbf{P}_{\mathrm{D}}(\omega)   \Big)
\end{eqnarray}
The detailed derivation can be found in Appendix~\ref{GMC-FRET-Simp}.

\subsection{Conventional MC-FRET in Near-field Region in Free Space}

The conventional MC-FRET\cite{silbey2,brumer} assumes that the short-distance near-field approximations in free space without any dielectric environment. Brumer and Coworkers extend FRET to the conventional MC-FRET with the classic approach\cite{brumer} in the free space. The generalized MC-FRET in Eq.~\ref{Qvectorfinal} can be reduced to the conventional MC-FRET in the free space. Since the DGF of the oscillating electric dipole in free space,
\begin{equation}\label{G0}
    \hat{\mathbf{G}}(\mathbf{r},\mathbf{r}_{0}, \mathit{k})=\frac{\exp(i\mathit{k}\mathit{R})}{4\pi R} \left[ \left(1+\frac{i\mathit{k}\mathit{R}-1}{\mathit{k}^2\mathit{R}^2}  \right)  \hat{\mathbf{I}} + \frac{3-3i \mathit{k} \mathit{R}-\mathit{k}^2\mathit{R}^2}{\mathit{k}^2\mathit{R}^2} \frac{\mathbf{R}\mathbf{R}}{\mathit{R}^2}  \right] ,
\end{equation}
where $\hat{\mathbf{I}}$ is the unit dyad, $\mathbf{R}=\mathbf{r}-\mathbf{r}_{0}$ is the vector between $\mathbf{r}$ and $\mathbf{r}_{0}$, $\mathit{R}$ is the distance, $\mathbf{R}\mathbf{R}$ is the outer product of $\mathbf{R}$, and  $\mathit{k}=\omega/c = \omega \sqrt{\epsilon_0 \mu_0}$, the
DGF in free space $\hat{\mathbf{G}}(\mathbf{r},\mathbf{r}_{0}, \mathit{k})$ is reduced to the electrostatic dipole dipole interaction without $\omega$ dependence in the near-field region according to the $1/R$ dominant term,
$\hat{\mathbf{G}}_{NF} (\mathbf{r},\mathbf{r}_{0}) \approx \frac{3\mathbf{n}\mathbf{n}-\hat{\mathbf{I}}}{4\pi \mathit{\omega}^2 \epsilon_0 \mu_0 \mathit{R}^3}$ where $\mathbf{n}=\frac{\mathbf{R}}{\mathit{R}}$. 

$\hat{\mathbf{G}}^{\mathrm{A}}$ and $\hat{\mathbf{G}}^{\mathrm{AD}}$ in the near-field region are reduced to electrostatic dipole-dipole interaction \cite{silbey2}. Therefore,
$ \textbf{Im}    \hat{\mathbf{G}}^{\mathrm{A}*} (\omega) \cdot \mathbf{P}_{\mathrm{A}}^*(\omega) \cdot \mathbf{P}_{\mathrm{A}}(\omega)$ in Eq.~\ref{Qvector} disappears. The generalized MC-FRET recovers the form of the conventional MC-FRET based on the electrostatic dipole dipole interaction\cite{brumer},
\begin{equation}
    \gamma_{MC}^{NF}
    = 2 \mu_0   \int_{0}^{\infty} d\omega \omega^3  \; \textbf{Im} \left(  \hat{\mathbf{G}}_{NF}^{\mathrm{AD}}  \cdot \mathbf{P}_{\mathrm{D}}^*(\omega)
    \cdot \mathbf{K}_{NF}(\omega)^{-1} \cdot \hat{\mathbf{G}}_{NF}^{\mathrm{AD}} \cdot \mathbf{P}_{\mathrm{D}}(\omega)   \right),
\end{equation}
where $\mathbf{K}_{NF}(\omega)={( \frac{c^2} {{\boldsymbol{\chi}}_A\omega^2}-\hat{\mathbf{G}}_{NF}^{\mathrm{A}}})$.
By defining $I_{kk'}^{\mathrm{A}}(\omega)=\omega^3 \textbf{Im} \big(\mathbf{K}_{NF}(\omega)^{-1}\big)_{kk'}$, $E_{jj'}^{\mathrm{D}}(\omega)= \mathbf{P}_{\mathrm{D}j}(\omega) \mathbf{P}_{\mathrm{D}j'}^*(\omega)$, $\gamma_{MC}$ in Eq.~\ref{Qvectorfinal} can be further reformatted to be the conventional MC-FRET rate\cite{silbey1},
\begin{equation}
    \gamma_{MC}^{NF} = \sum_{kk',jj'} \hat{\mathbf{G}}_{NF,kj}^{\mathrm{AD}} \hat{\mathbf{G}}_{NF, k'j'}^{\mathrm{AD}}  \int_0^\infty d\omega I_{kk'}^{\mathrm{A}}(\omega) E_{jj'}^{\mathrm{D}}(\omega).
\end{equation}

\section{MC-FRET near Metallic Thin Films}\label{MCS}
\begin{figure}[H]
    \centering
    \includegraphics[width=0.9\textwidth]{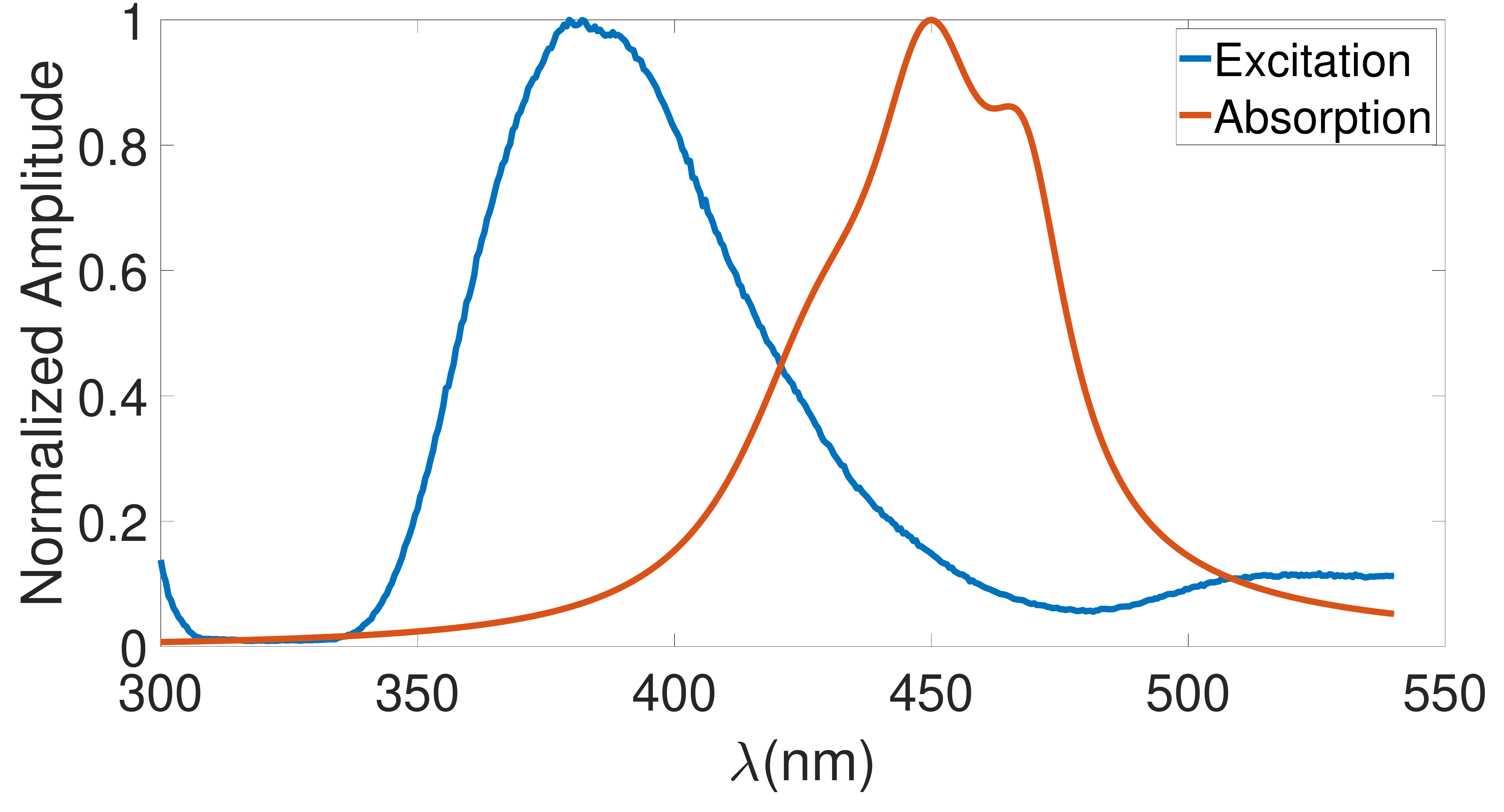}
    \caption{MCA donor emission spectrum (blue line) and C6 acceptor absorption spectrum (orange). Both spectra are normalized}\label{spectrum}
\end{figure}

The MC systems in the vicinity of the non-equilibrium evanescent near field will strongly modulate MC-FRET. Two artificial MC systems are studied above the metal thin film surface. One is the one-to-two MC structure and the other is three-fold ring-ring structure. In the artificial MC systems, the donor chromophore is 7-methoxycoumarin-4-acetic acid (MCA), and the acceptor chromophore is coumarin 6 (C6)\cite{dixon2005photochemcad,farinotti19834,reynolds1975new}. The MCA emission spectrum and the C6 absorption spectrum are shown in Fig.~\ref{spectrum}. The largest overlap of the MCA emission and C6 absorption spectra are located around 420 nm$^{-1}$ wave length.

The induced polarizability $\chi_A(\omega)$ of the acceptor chromophore molecule is needed. The imaginary part of the induced polarizability $\chi(\omega)$ can be derived from the absorption cross-section\cite{darby2016modified}, $\sigma_{\mathrm{abs}}(\omega)$ as,
\begin{equation}
    \operatorname{Im}\left(\chi(\omega)\right)=\frac{9 \epsilon_{0} \sqrt{\epsilon_{M}}}{\left(\epsilon_{M}+2\right)^{2}} \frac{\lambda}{2 \pi} \sigma_{\mathrm{abs}}(\lambda),
\end{equation}
where $\epsilon_{M}$ is the dielectric constant of the medium and in ethanol, $\epsilon_{M}$=25.8, and $\epsilon_{0}$ the permittivity of free space. The Kramers-Kr$\ddot{o}$nig relation gives the expression of the induced polarizability as the function of the wavelength $\lambda=(2\pi c)/\omega$,
\begin{equation}\label{chi}\chi(\lambda)=\alpha_{\text {static }}+\sum_{n=0,1,2} \frac{\alpha_{n} \lambda_{n}}{\mu_{n}}\left[\frac{1}{1-\frac{\lambda_{n}^{2}}{\lambda^2}-i \frac{\lambda_{n}^{2}}{\lambda \mu_{n}}}-1\right]\end{equation},
where
$\lambda_{0}=427.8908 \mathrm{nm} \; \mathrm{S.I.}$,  $\mu_{0}=4762.6 \mathrm{nm} \; \mathrm{S.I.}$,  $\alpha_{0} =3.8104 \times 10^{-36}\; \mathrm{S.I.}$, $\lambda_{1}=468.2834\; \mathrm{nm}$,  $\mu_{1}=10912\; \mathrm{nm}$,  $\alpha_{1} =5.0689 \times 10^{-36}\; \mathrm{S.I}$, $\lambda_{2}=449.5650 \mathrm{nm} \; \mathrm{S.I.}$,  $\mu_{2}=7299.3 \mathrm{nm} \; \mathrm{S.I.}$,  $\alpha_{2} =7.6581 \times 10^{-36}\; \mathrm{S.I.}$, and $\alpha_{\text {static }} =5.9091 \times 10^{-39} \; \mathrm{S.I}$
The classic DGF is evaluated for the study of the coupling factor with MC-FRET in the evanescent EV wave about the metal thin film. The scattering DGF of the metal thin film and hyperbolic multi-layer thin film has the analytical expression. The detailed derivation of the scattering DGF is presented in Appendix~\ref{ADGF}.

\subsection{Distance Dependence}
\begin{figure}[H]
    \centering
    \includegraphics[width=0.9\textwidth]{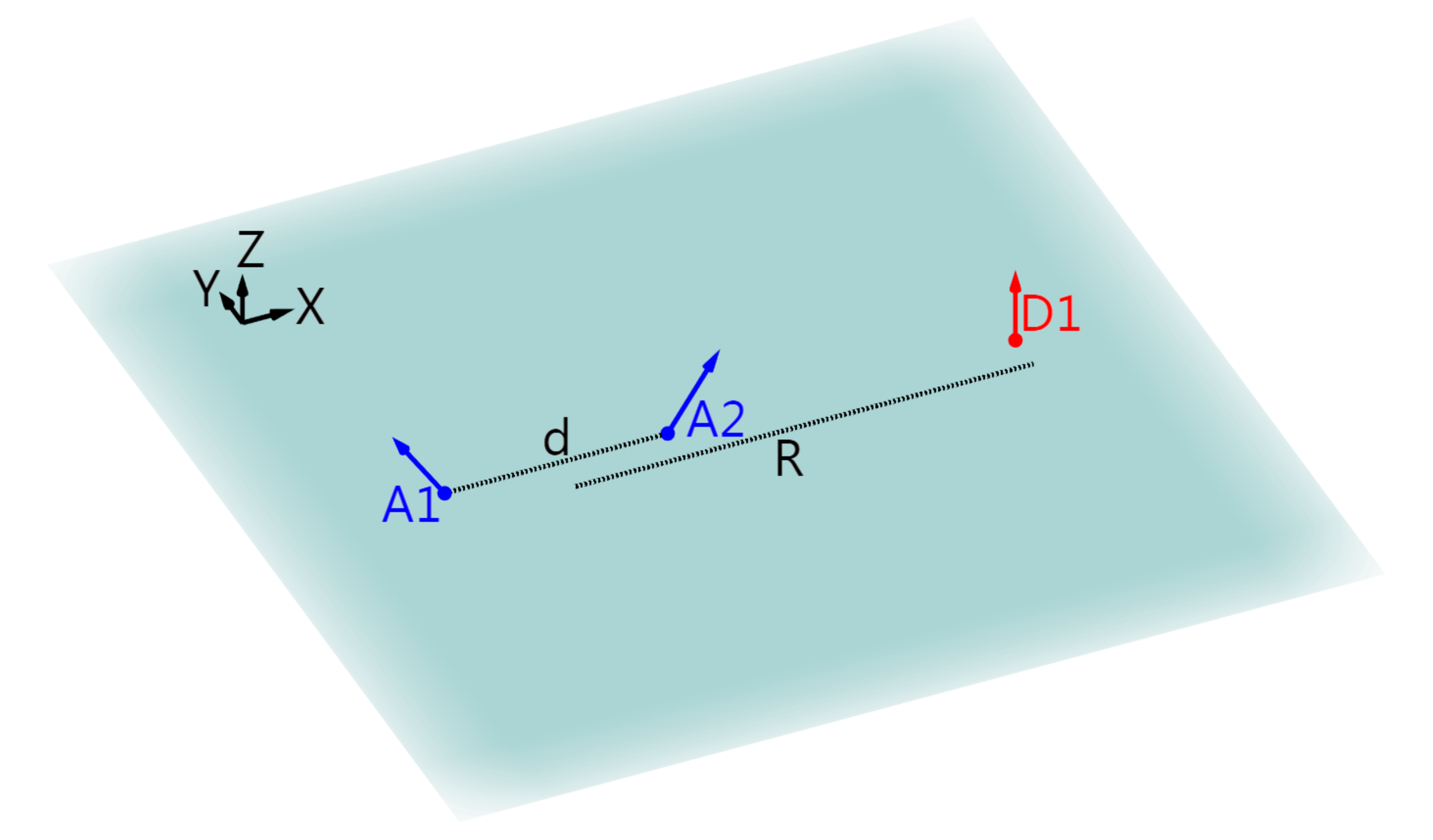}
    \caption{The setup shows the one-to-two MC structure where the donor is located at $(0,R,5nm)$ and two acceptor $A_1$ and $A_2$ are located at $(0,-0.5{\times}d,5nm)$ and $(0,0.5{\times}d,5nm)$. d is the distance between the two acceptors and R the distance from the donor to the center of the two acceptors. The polarization directions of A1, A2 and D1 are (-1, 0, 1) (1, 0, 1) and (0, 0, 1) respectively.}\label{1to2sketch}
\end{figure}

\begin{figure}[H]
    \centering
    \subfigure[d = 1 nm]{\includegraphics[width=0.496\textwidth]{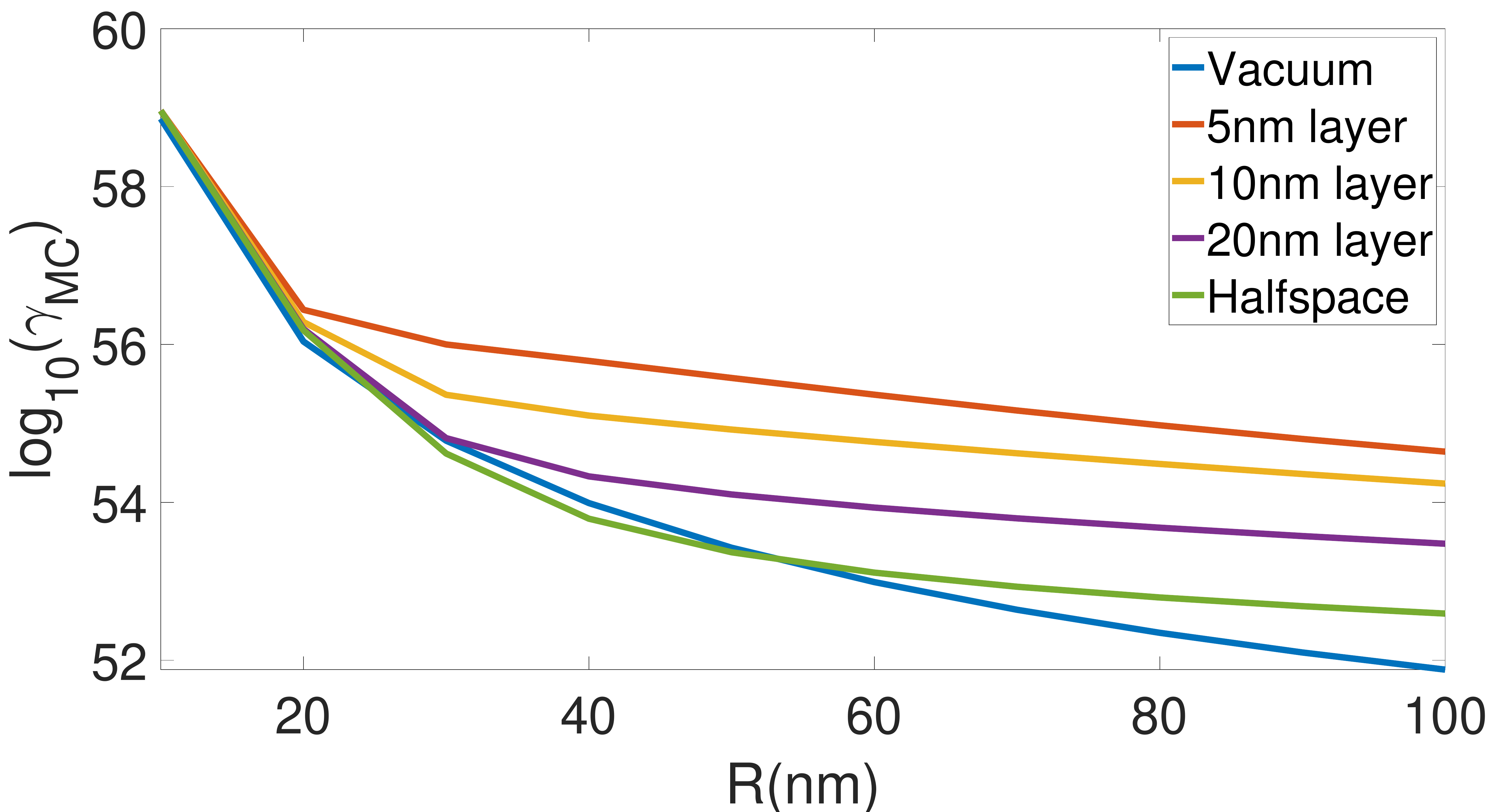}}\label{compare_1to2_d=1}
    \subfigure[d = 1 nm]{\includegraphics[width=0.496\textwidth]{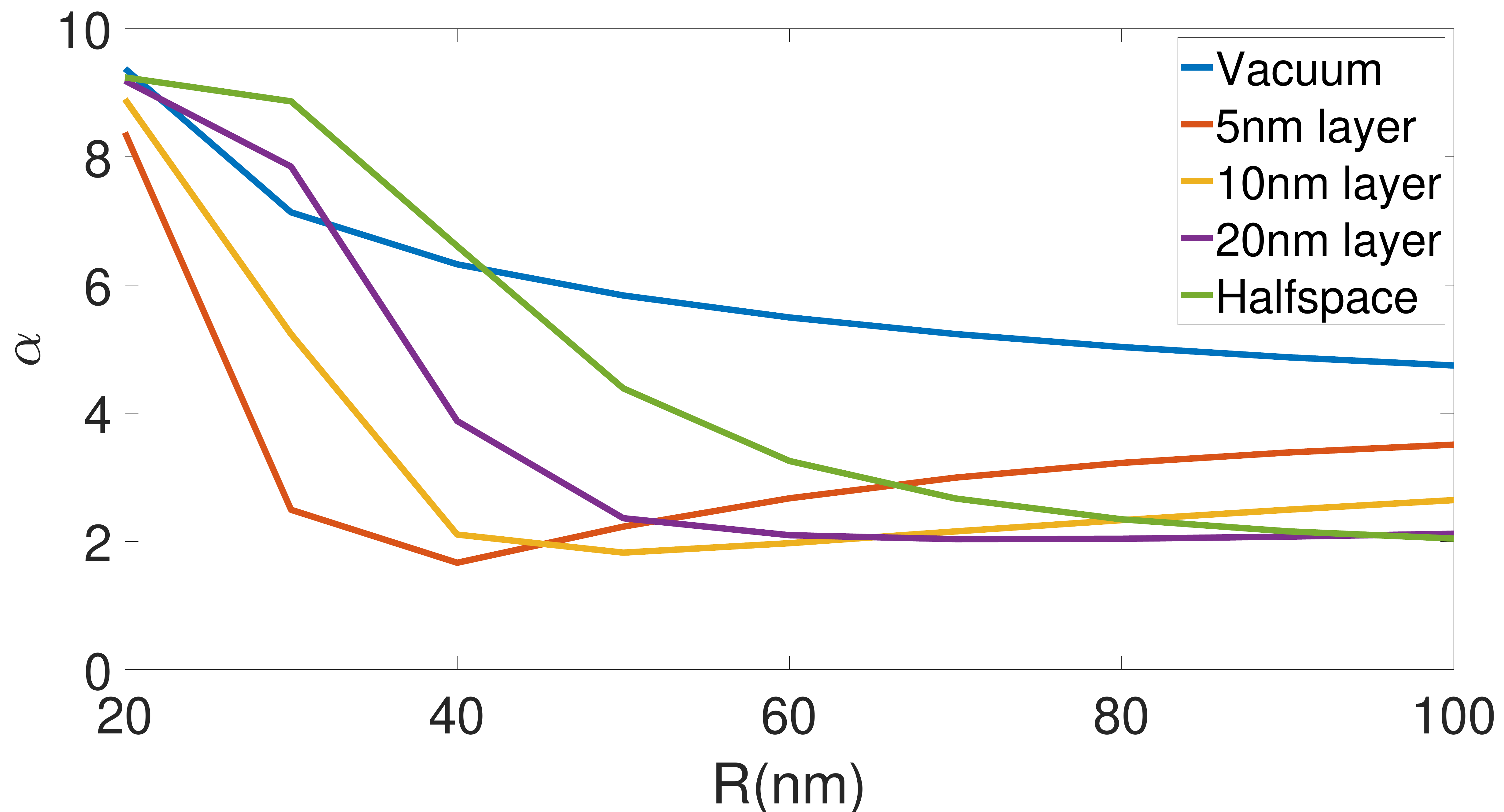}}\label{alpha1}\\
    \subfigure[d = 5 nm]{\includegraphics[width=0.496\textwidth]{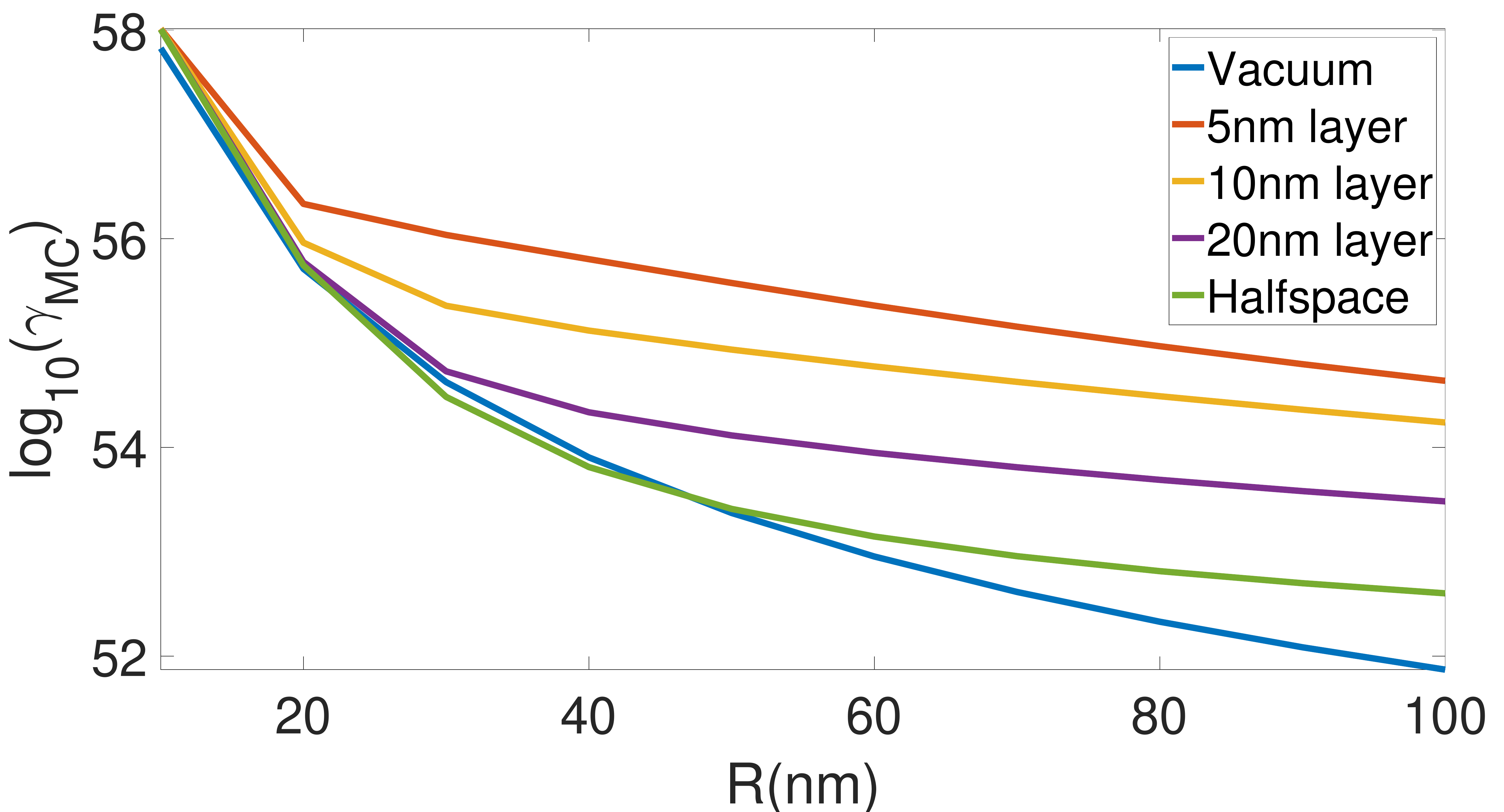}}\label{compare_1to2_d=5}
    \subfigure[ d = 5 nm]{\includegraphics[width=0.496\textwidth]{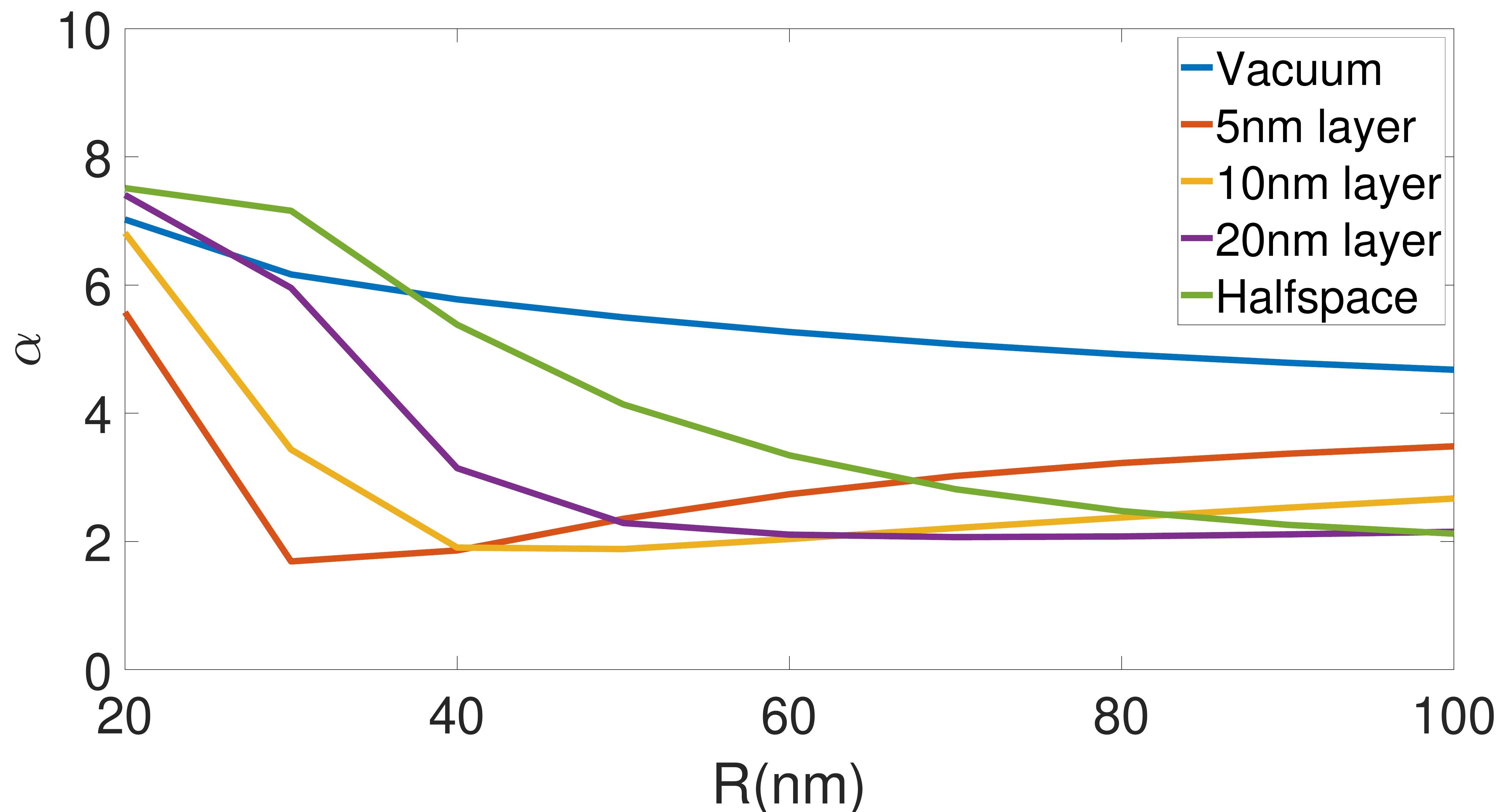}}\label{alpha5}\\
    \subfigure[d = 10 nm]{\includegraphics[width=0.496\textwidth]{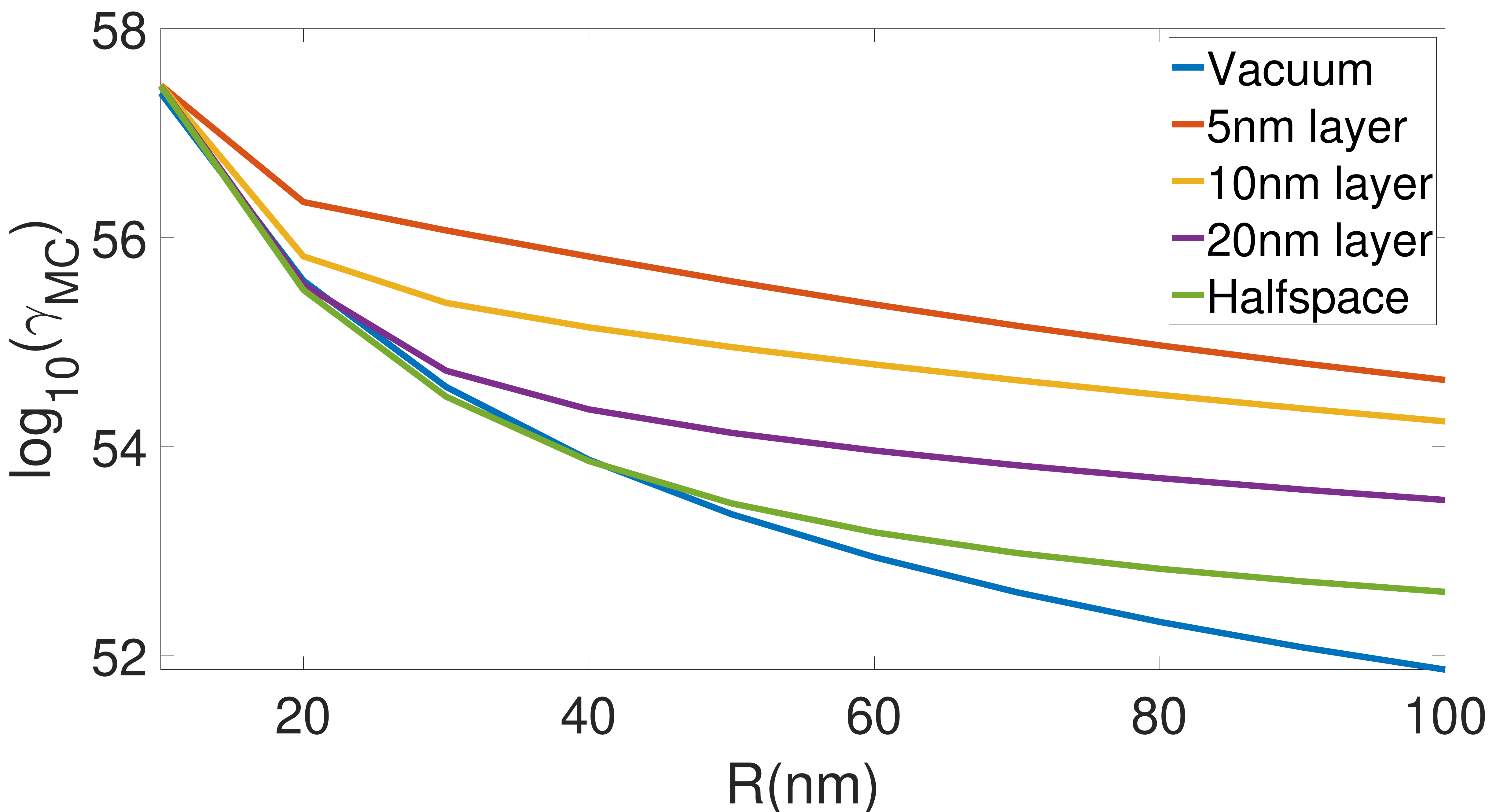}}\label{compare_1to2_d=10}
    \subfigure[ d = 10 nm]{\includegraphics[width=0.496\textwidth]{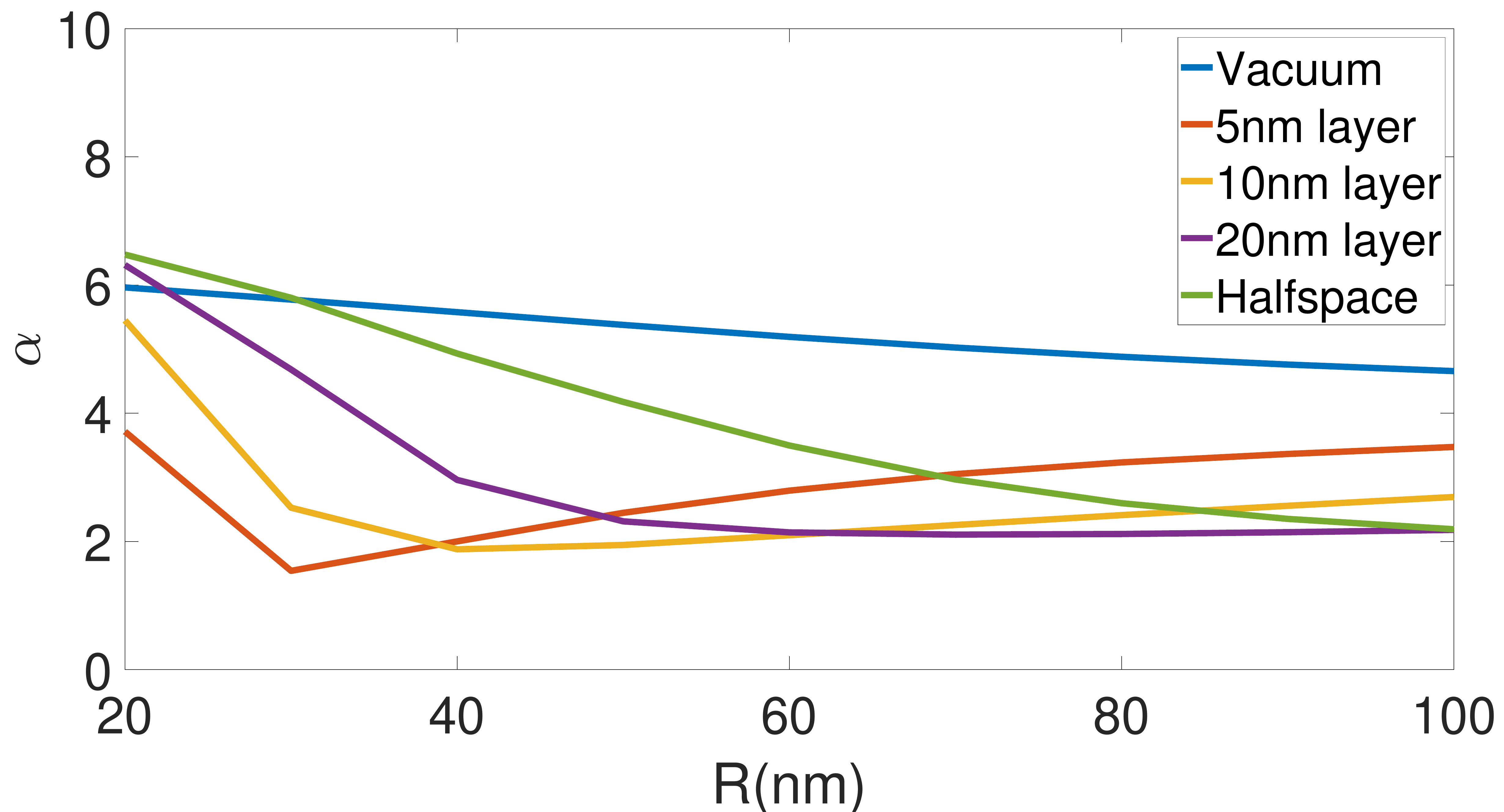}}\label{alpha10}\\
    \caption{The MC-FRET rates $\gamma_{MC}$ and $\alpha$ above the four different silver thin metallic films, infinite halfspace, 5 nm, 10 nm and 20 nm single-layer thin films.}\label{compare_1to2}
\end{figure}

Two MC structures, the one-to-two MC structure and three-fold ring-to-ring MC structure, are used to study the MC-FRET distance dependence in the vicinity of metallic thin film. The one-to-two MC structure includes one donor and two acceptors. The geometry of the one-to-two MC structure is shown in Fig.~\ref{1to2sketch}. The one-to-two MC structure has two geometric factors, the separation distance $d$ between the two acceptors  and the distance $R$ between the center of the two acceptors and the donor. 
The modulation of MC-FRET in the vicinity of four metallic surfaces, the silver infinite halfspace, 5nm, 10nm and 20nm single-layer silver thin films, are studied in reference to the vacuum. The MC-FRET rate $\gamma_{MC}$ shows very different R distance-dependence for different d. To quantitatively describes R distance dependence in the MC systems, the coefficient $\alpha$ is defined as,
\begin{equation}
    \alpha=-\frac{d\ln(\gamma_{MC})}{d\ln(R)}
\end{equation}
$\gamma_{MC}$ and $\alpha$ with different R and d are shown in Fig.~\ref{compare_1to2}.

When the separation distance $d$, {\it{i.e.}} the two acceptors are far from each other and R is extremely small, the MC-FRET rate $\gamma_{MC}$ distance dependence approximately approaches the conventional FRET $R^{-6}$ distance-dependence with $\alpha=6$.  MC-FRET can be reduced into conventional FRET. $\alpha$ generally decreases as R increases. The metallic interface can slow down the decay of $\gamma_{MC}$ with R as shown in Fig.~\ref{compare_1to2}. When the separation distance d between the two acceptors increases from 1 nm to 10 nm, the MC-FRET rates decay slower due to the weakening of interactions within the acceptor aggregator. 

\begin{figure}[H]
    \centering
    \subfigure[The setup of the three-fold ring-to-ring MC structure above metal surface. All the molecules are polarized perpendicular to the surface.]{\includegraphics[width=0.9\textwidth]{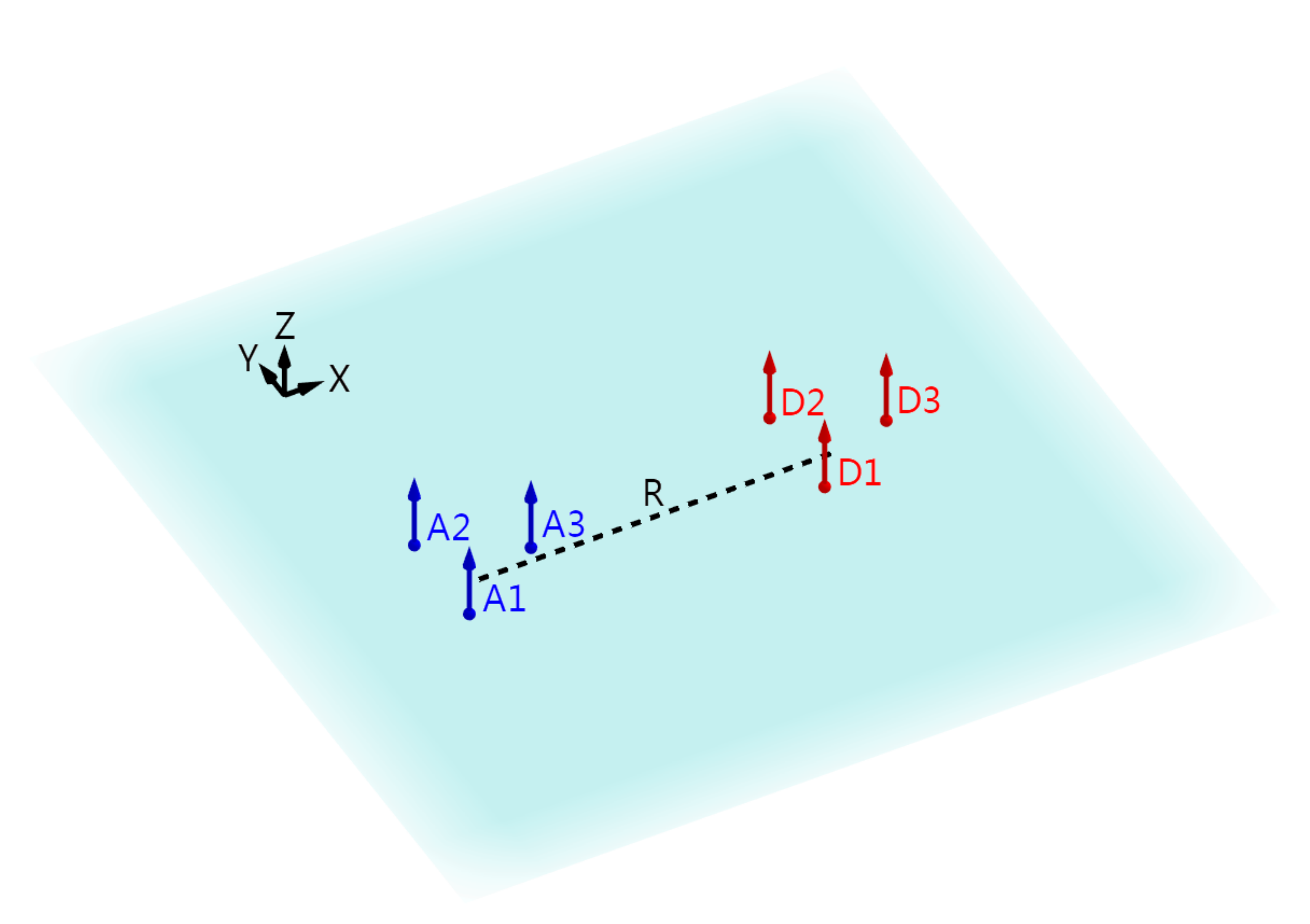}}\label{birdview} \\
    \subfigure[Rotating angles $\theta$ of the three-fold ring. The radius of the ring $R_0$ is 5nm]{\includegraphics[width=0.9\textwidth]{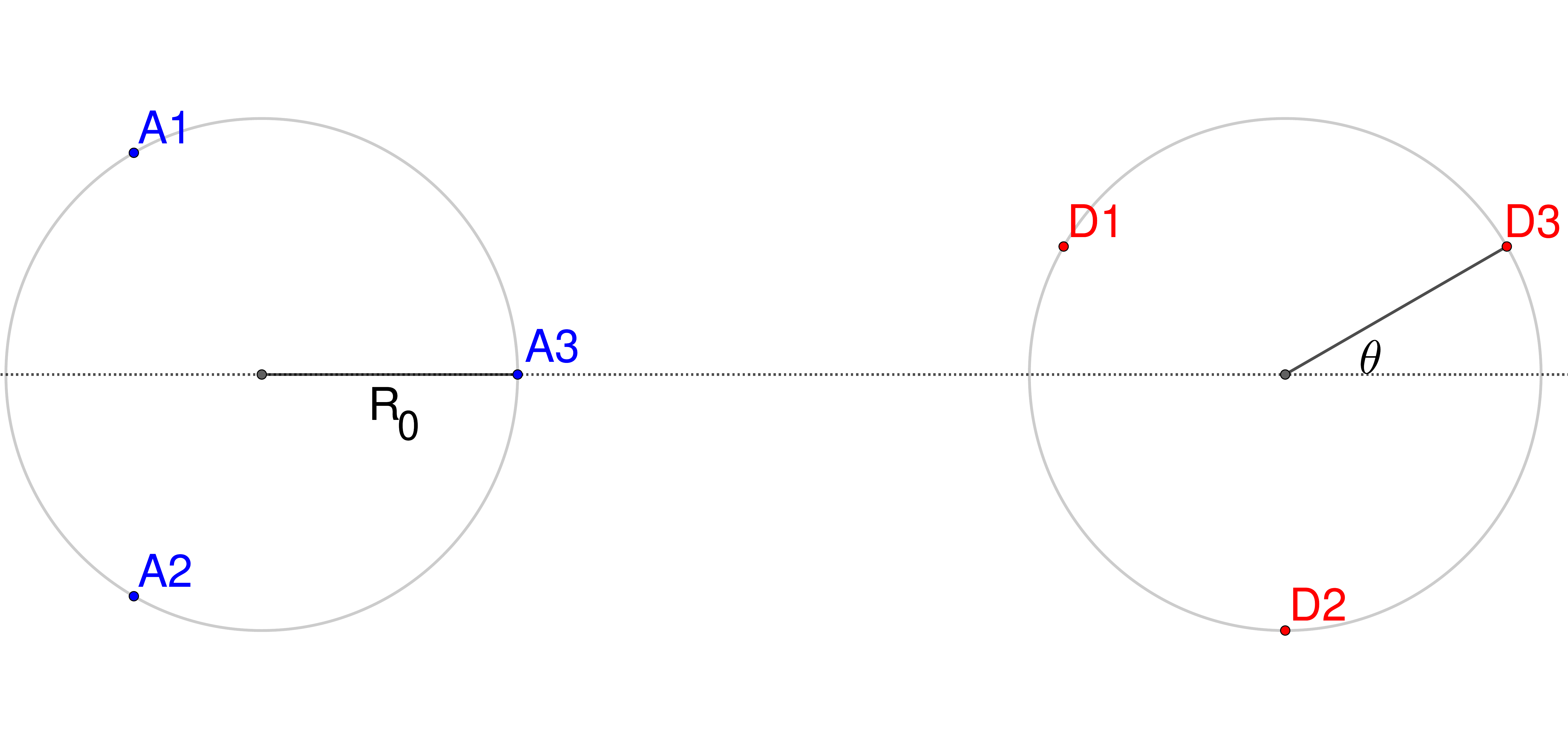}}\label{rotating}
    \caption{The three-fold ring-to-ring MC structure}\label{RingtoRingSketch}
\end{figure}

\begin{figure}[H]
    \centering
    \includegraphics[width=0.9\textwidth]{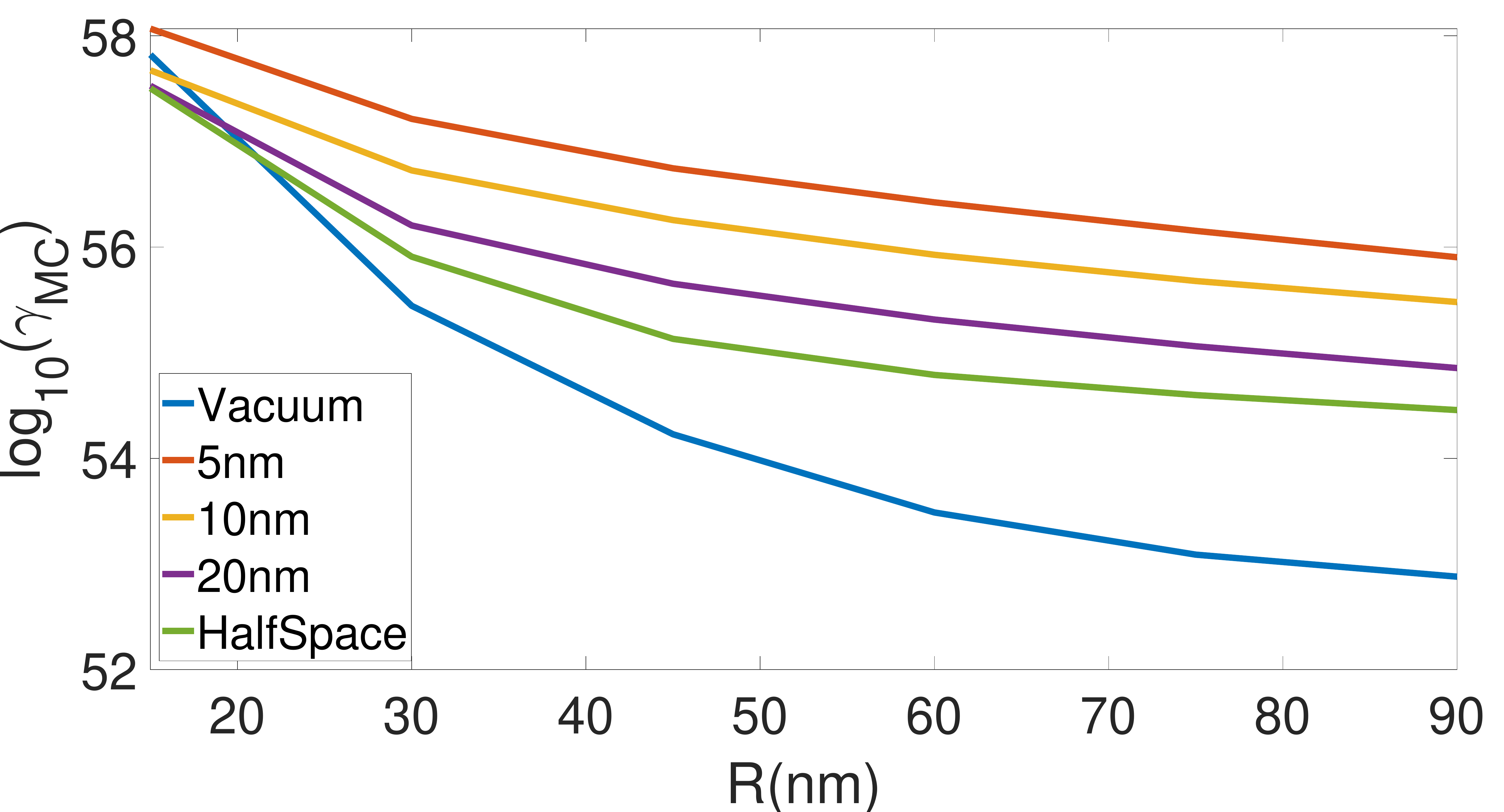}
    \caption{The MC-FRET rates of three-fold ring-to-ring MC system with different distance R in vacuum and above the four different thin metallic film, infinite halfspace, 5 nm, 10 nm and 20 nm single-layer thin films}\label{RingToRing2}
\end{figure}

\begin{figure}[H]
    \centering
    \subfigure[$R=15nm$]{\includegraphics[width=0.496\textwidth]{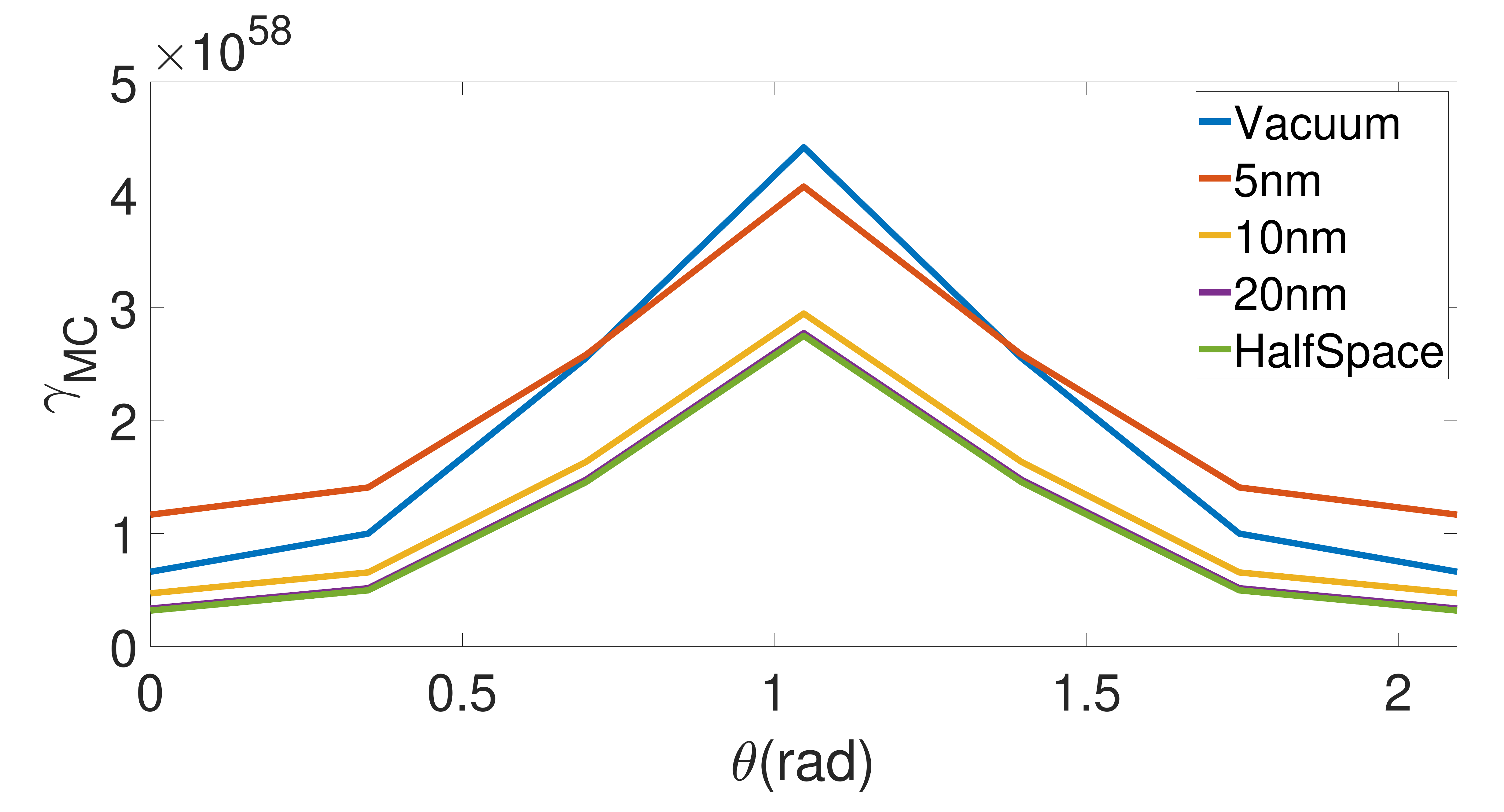}}\label{Ring15}
    \subfigure[$R=30nm$]{\includegraphics[width=0.496\textwidth]{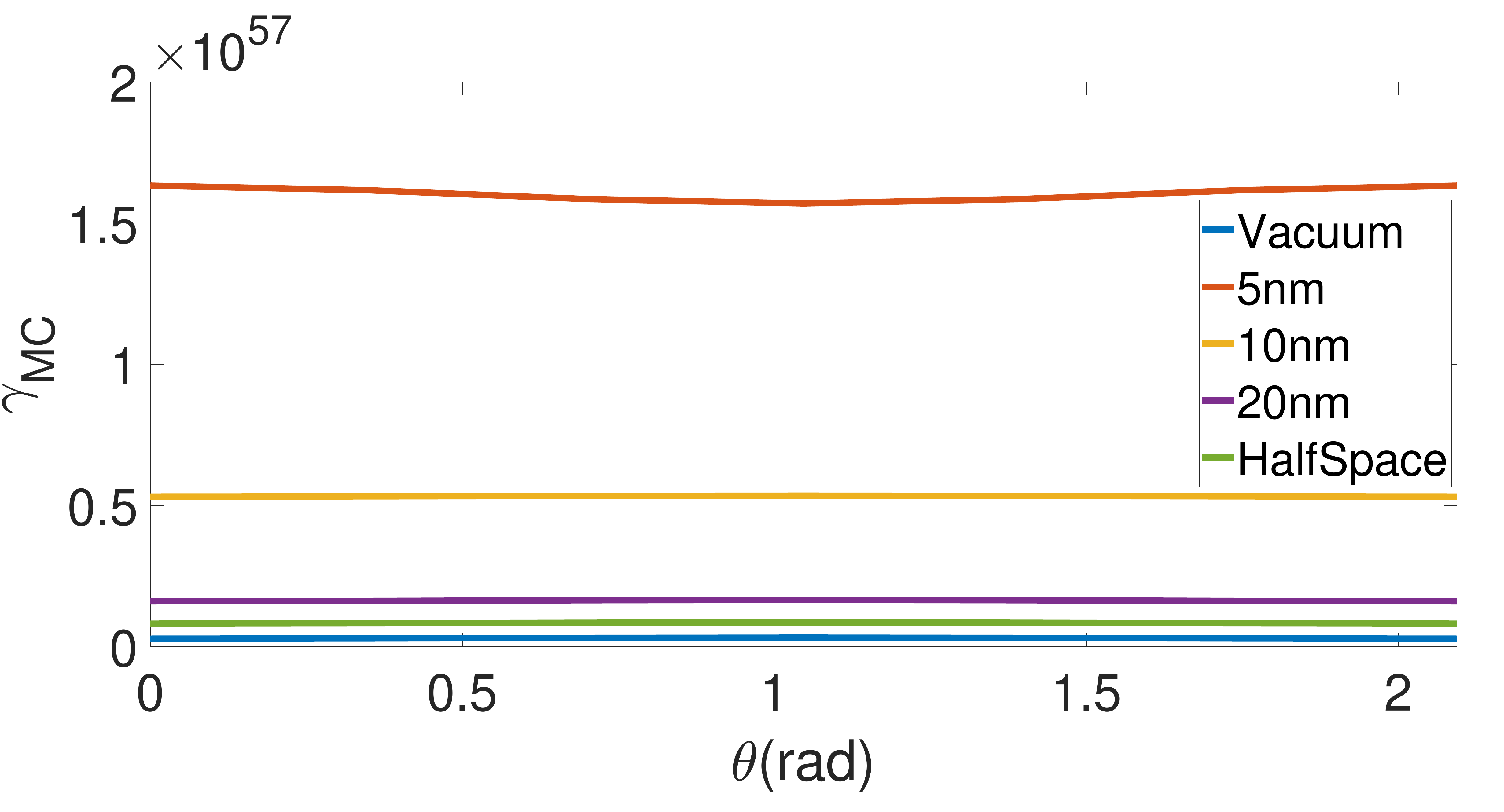}}\label{Ring30}%
    \caption{ (a) The MC-FRET rate $\gamma_{MC}$ for different rotating angles $\theta$ with R = 15 nm in vacuum and above four thin metallic films, (b) the MC-FRET rate $\gamma$ against rotating angle at R = 30 nm in vacuum and above the four thin metallic films. The four thin metallic films include infinite half-space, 5 nm, 10nm, 20nm single-layer thin film.}\label{RingtoRing}
\end{figure}

Furthermore, the ring structures with the N-fold symmetry widely exist photosynthetic light-harvesting complex\cite{cao1}. The three-fold ring-to-ring structure is the simplest ring structure. In the vicinity of the silver thin film, the structure of the three-fold ring-to-ring is shown in Fig.~\ref{RingtoRingSketch}. In the three-fold ring-to-ring structure, it has key geometric factor $\theta$ and R distance as shown in Fig.~\ref{RingtoRingSketch}.
In the three-fold ring-to-ring MC structure, the chromophores have the same induced polarizability in the donor and acceptor aggregates respectively. The calculation details of MC-FRET rate are given Appendix~\ref{3fold}. With $\theta=0$, Fig.~\ref{RingToRing2} shows the $R$ distance dependence of the MC-FRET rates of the three-fold ring-to-ring MC systems above the four different thin metallic films, infinite halfspace, 5nm, 10nm and 20nm single-layer silver thin films in reference to the vacuum. And the $\theta$ dependence of MC-FRET with $R=15nm$ and $R=30nm$ was shown in Figutre~\ref{RingtoRing}.

\section{Polarization Orientation Dependence}\label{mod}
\begin{figure}[H]
    \centering
    \subfigure[XX configuration]{\includegraphics[width=0.495\textwidth]{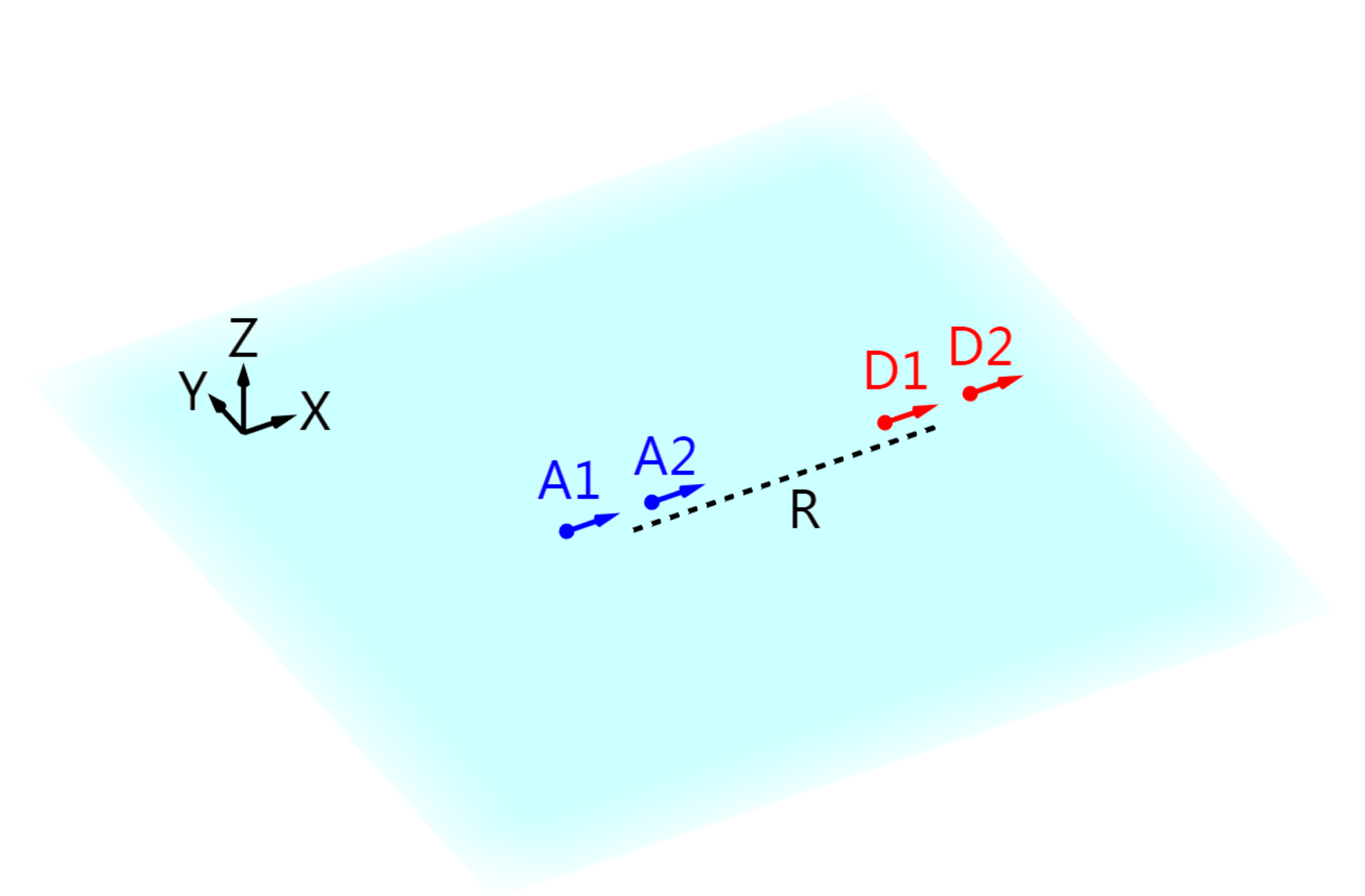}}\label{2to2x}
    \subfigure[YY configuration]{\includegraphics[width=0.495\textwidth]{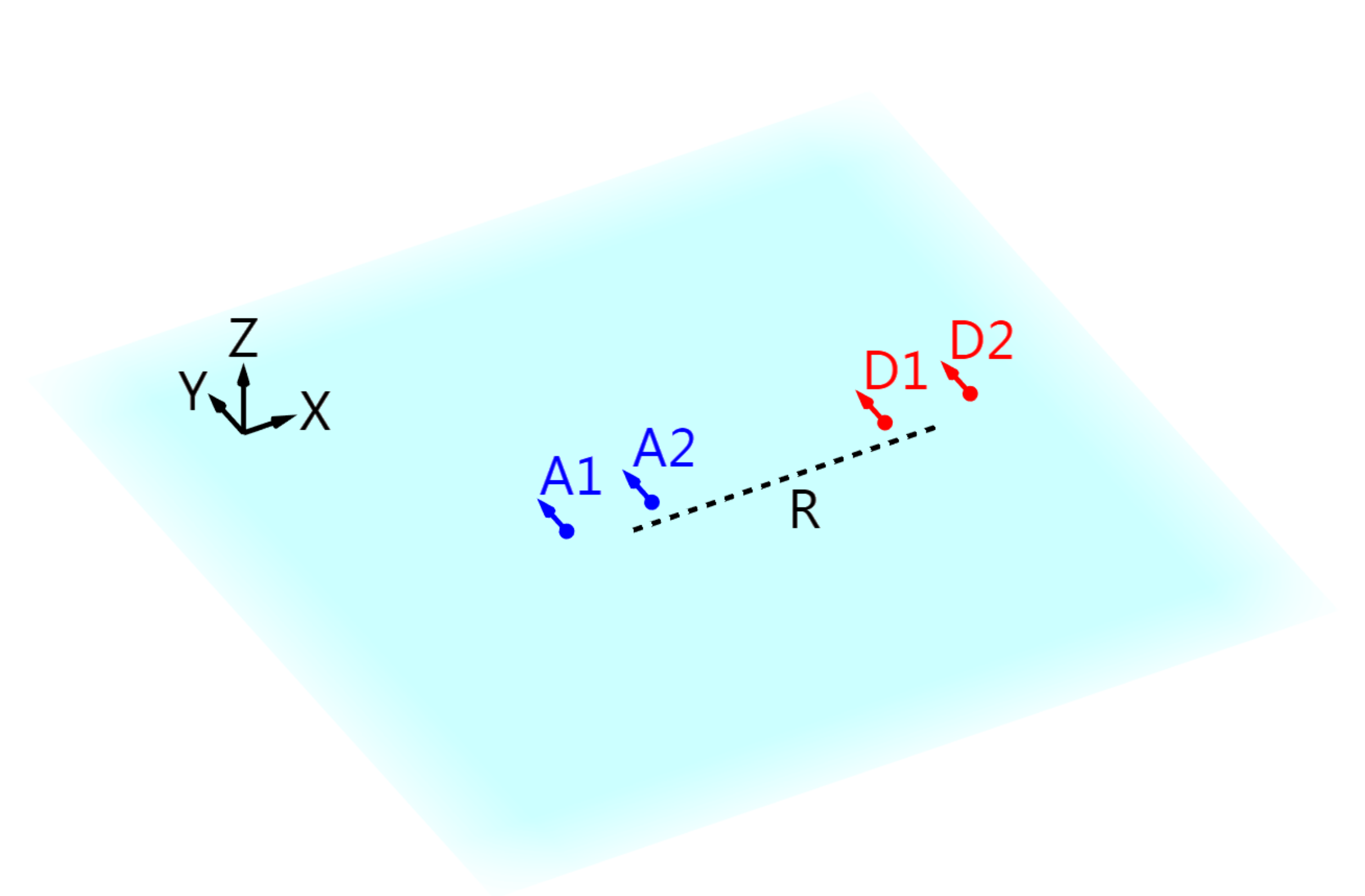}}\label{2to2y}\\
    \subfigure[ZZ configuration]{\includegraphics[width=0.495\textwidth]{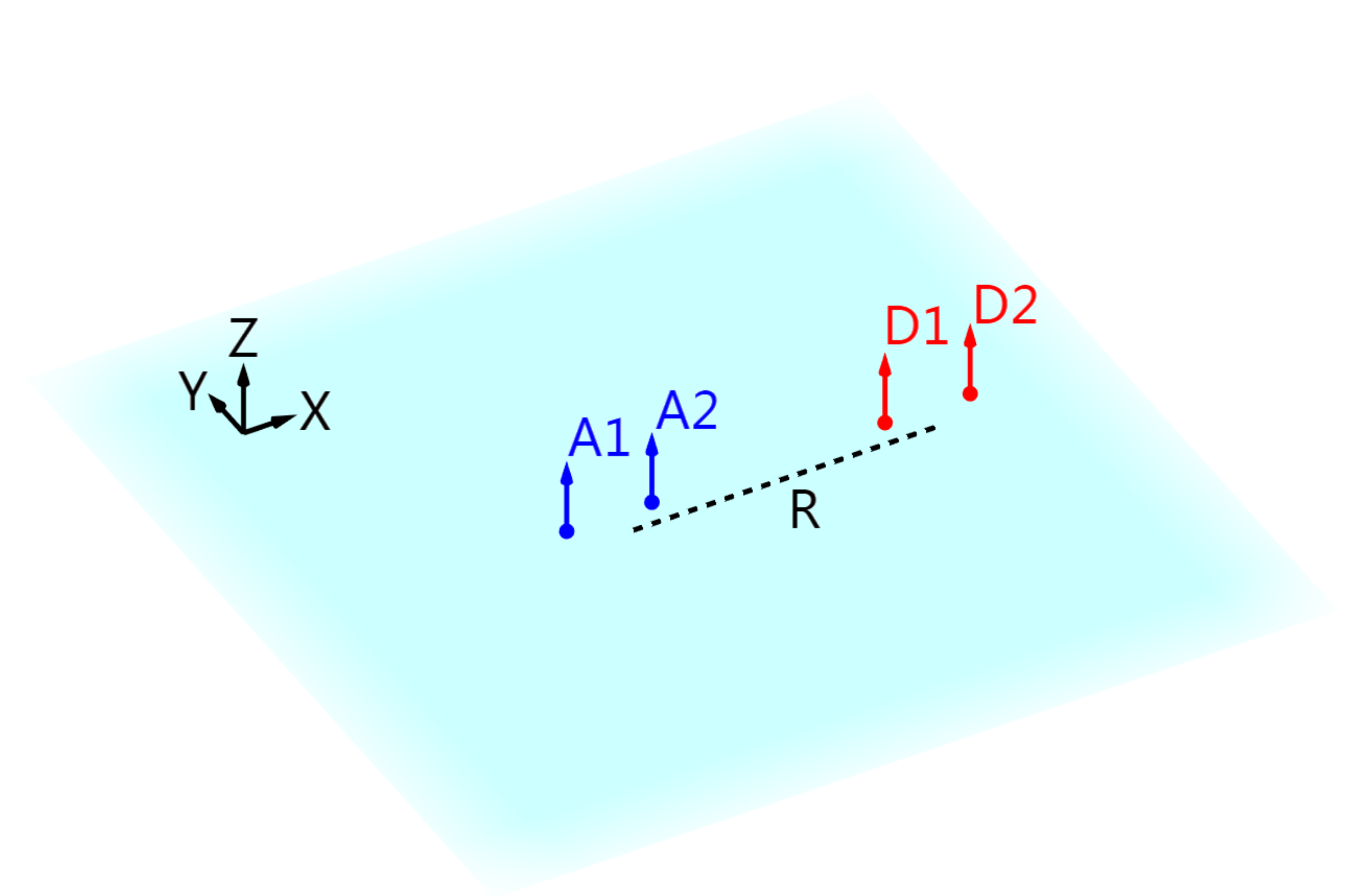}}\label{2to2z}%
    \caption{Two donor chromophores (D1, D2) and two acceptor chromophores (A1, A2) are located at 5 nm above thin metallic film, the four chromophores $A_1$, $A_2$, $D_1$ and $D_2$ are located at (0, 0, 5)nm,  (5, 0, 5)nm,  (R, 0, 5)nm, and  (R+5, 0, 5)nm respectively. The distance between the two acceptors is 5 nm. The distance between the two donors is also 5 nm. R is the distance between the centers of the donors and acceptors. The polarization directions of all molecules are set as X direction(a), Y direction(b) and Z direction(c) in turn, which are called XX, YY and ZZ configurations respectively.  }\label{2to2sketch}
\end{figure}

\begin{figure}[H]
    \centering
    \subfigure[$\gamma/\gamma_{0}$ in XX configuration ]{\includegraphics[width=0.496\textwidth]{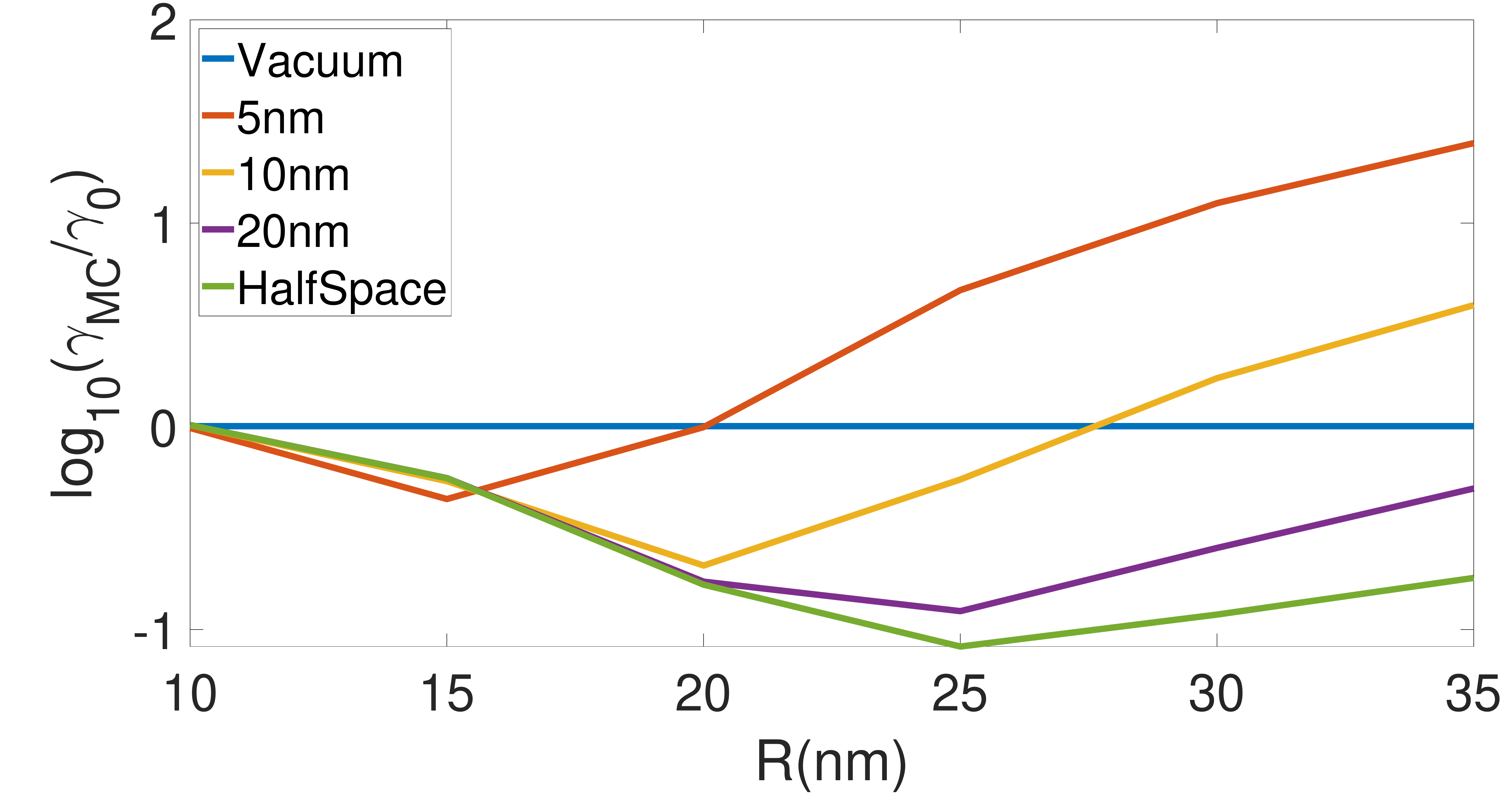}}\label{Gamma_x}
    \subfigure[$\beta_{xx}$]{\includegraphics[width=0.496\textwidth]{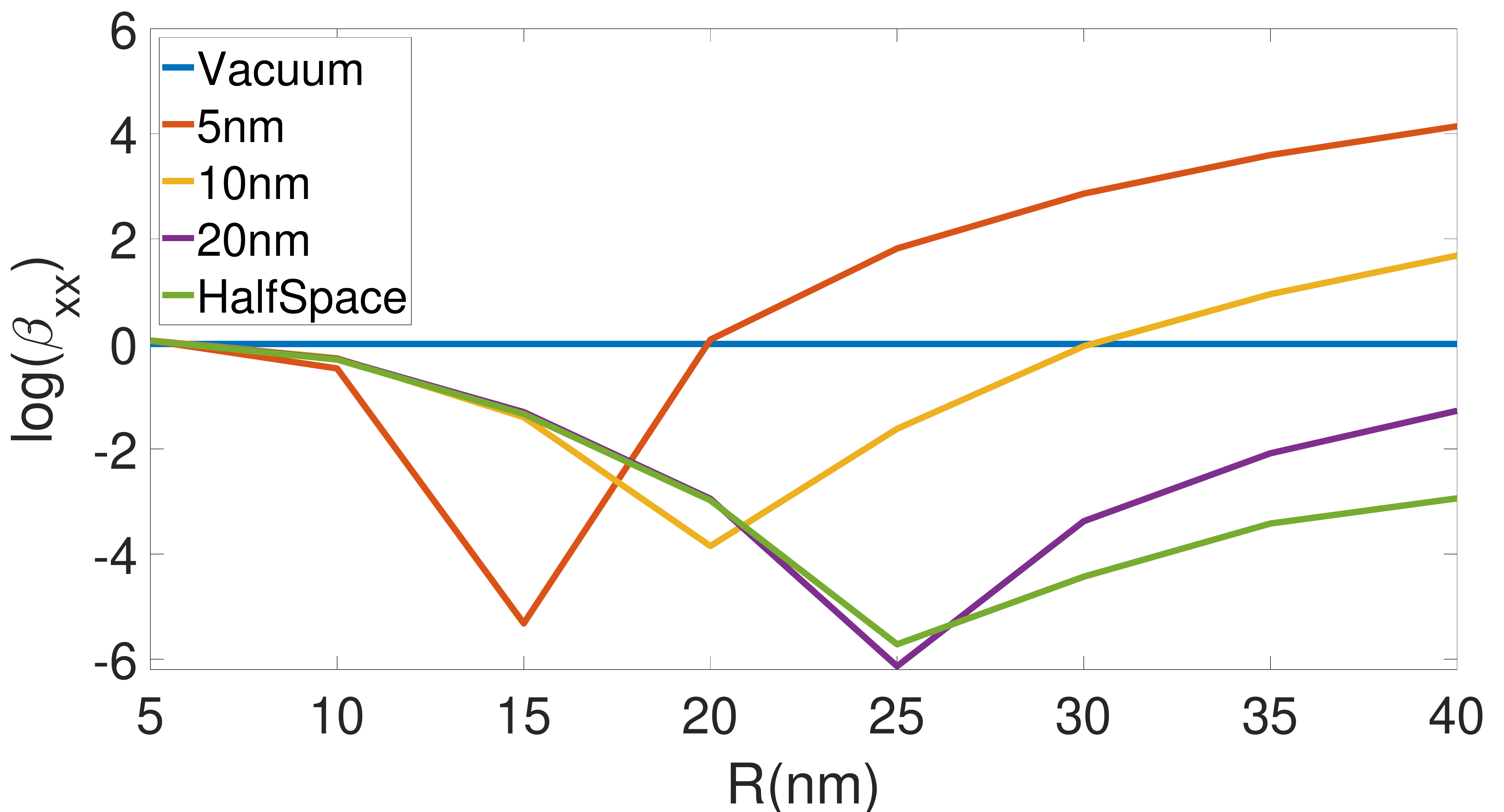}}\label{Beta_x}\\
    \subfigure[$\gamma/\gamma_{0}$ in YY configuration]{\includegraphics[width=0.496\textwidth]{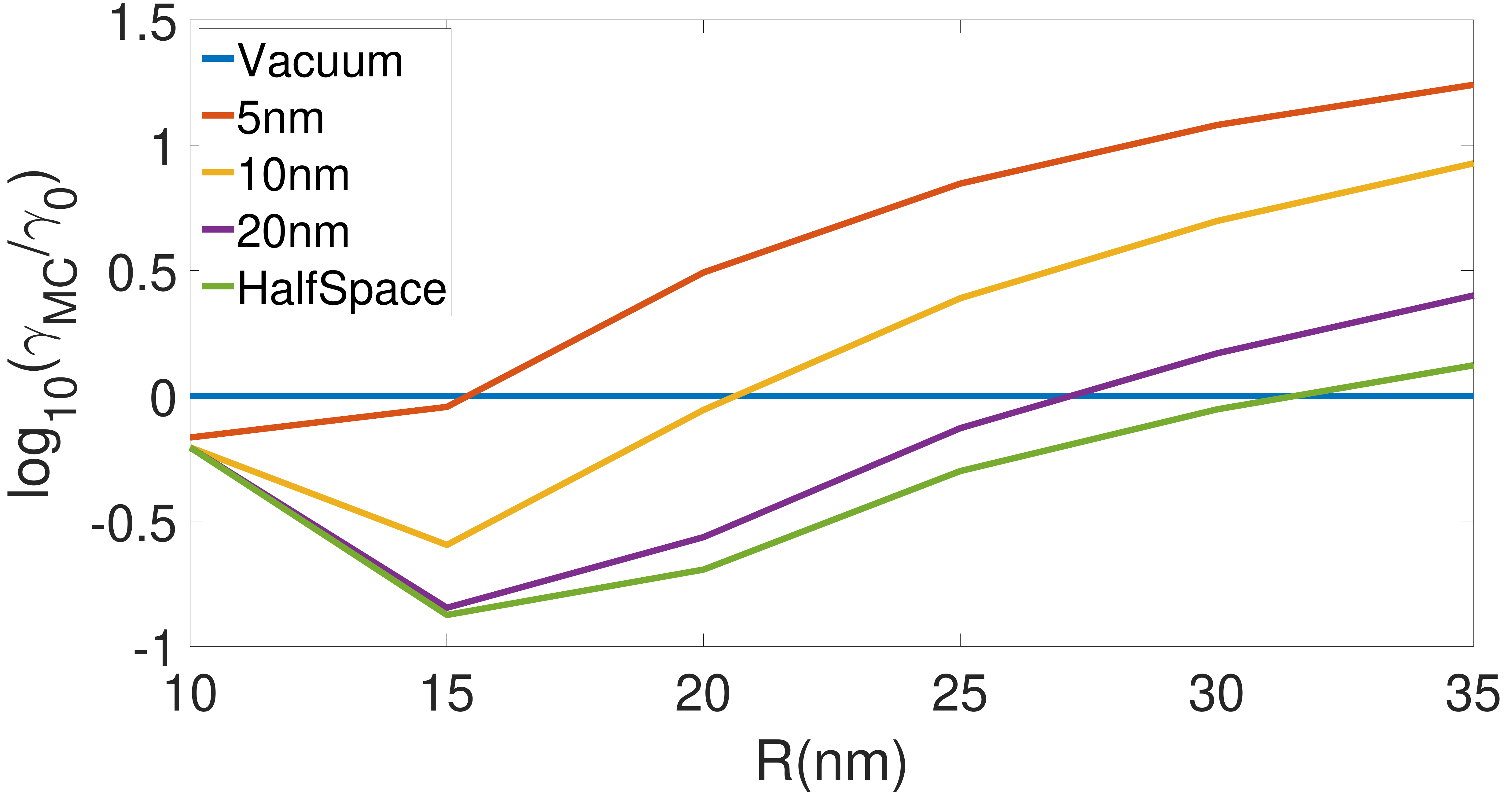}}\label{Gamma_y}
    \subfigure[$\beta_{yy}$]{\includegraphics[width=0.496\textwidth]{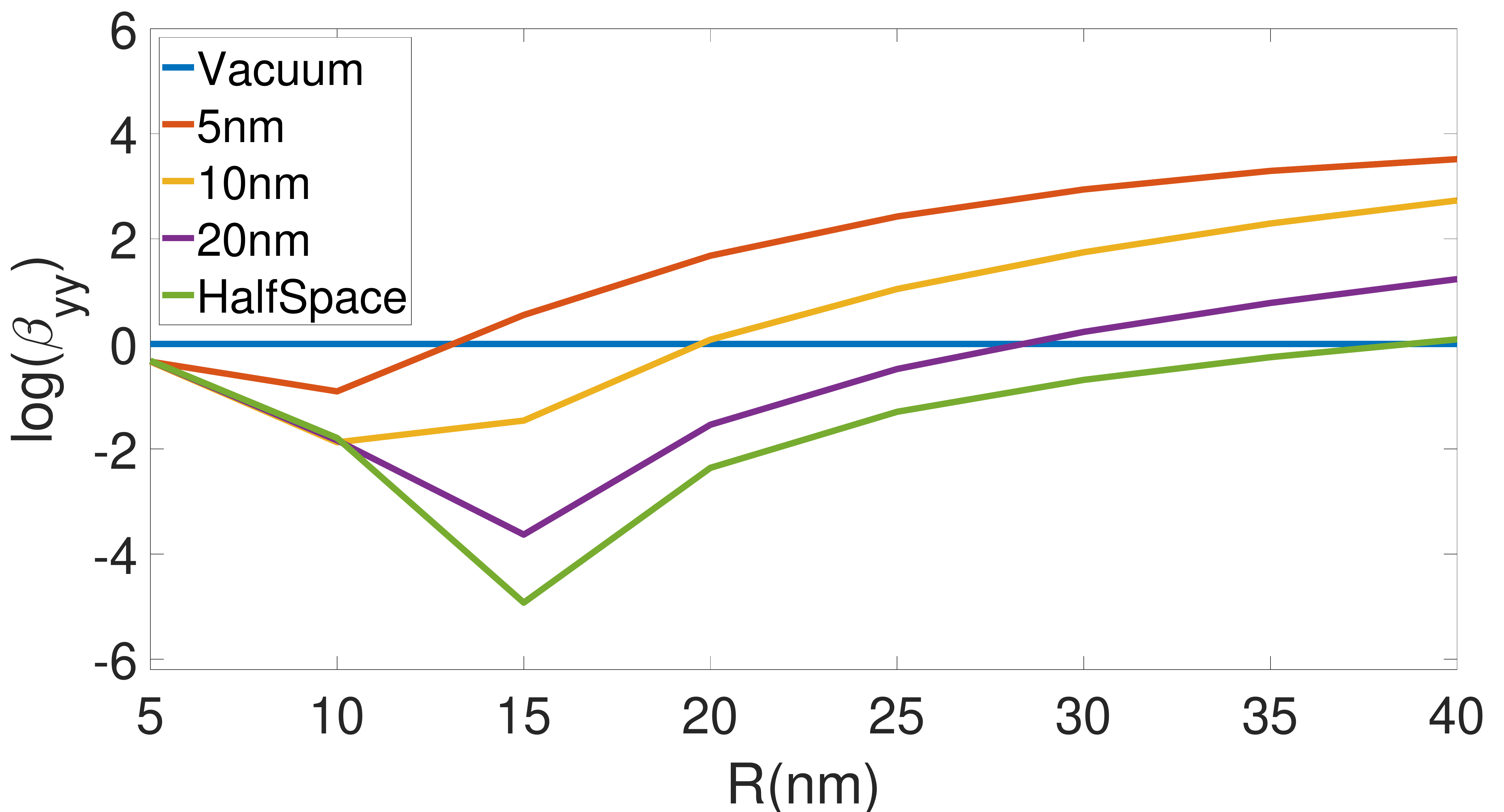}}\label{Beta_y}\\
    \subfigure[$\gamma/\gamma_{0}$ in ZZ configuration]{\includegraphics[width=0.496\textwidth]{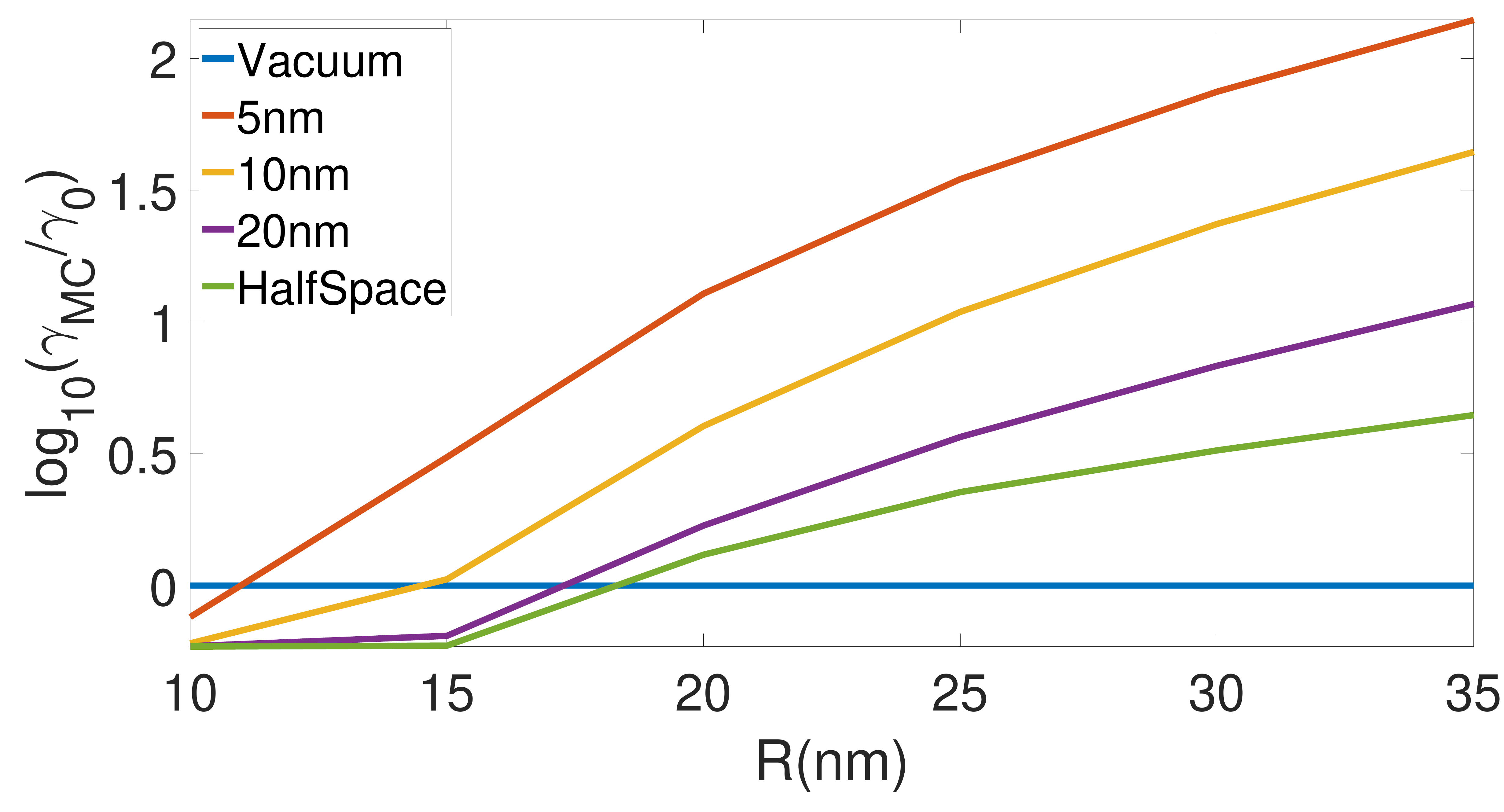}}\label{Gamma_z}
    \subfigure[$\beta_{zz}$]{\includegraphics[width=0.496\textwidth]{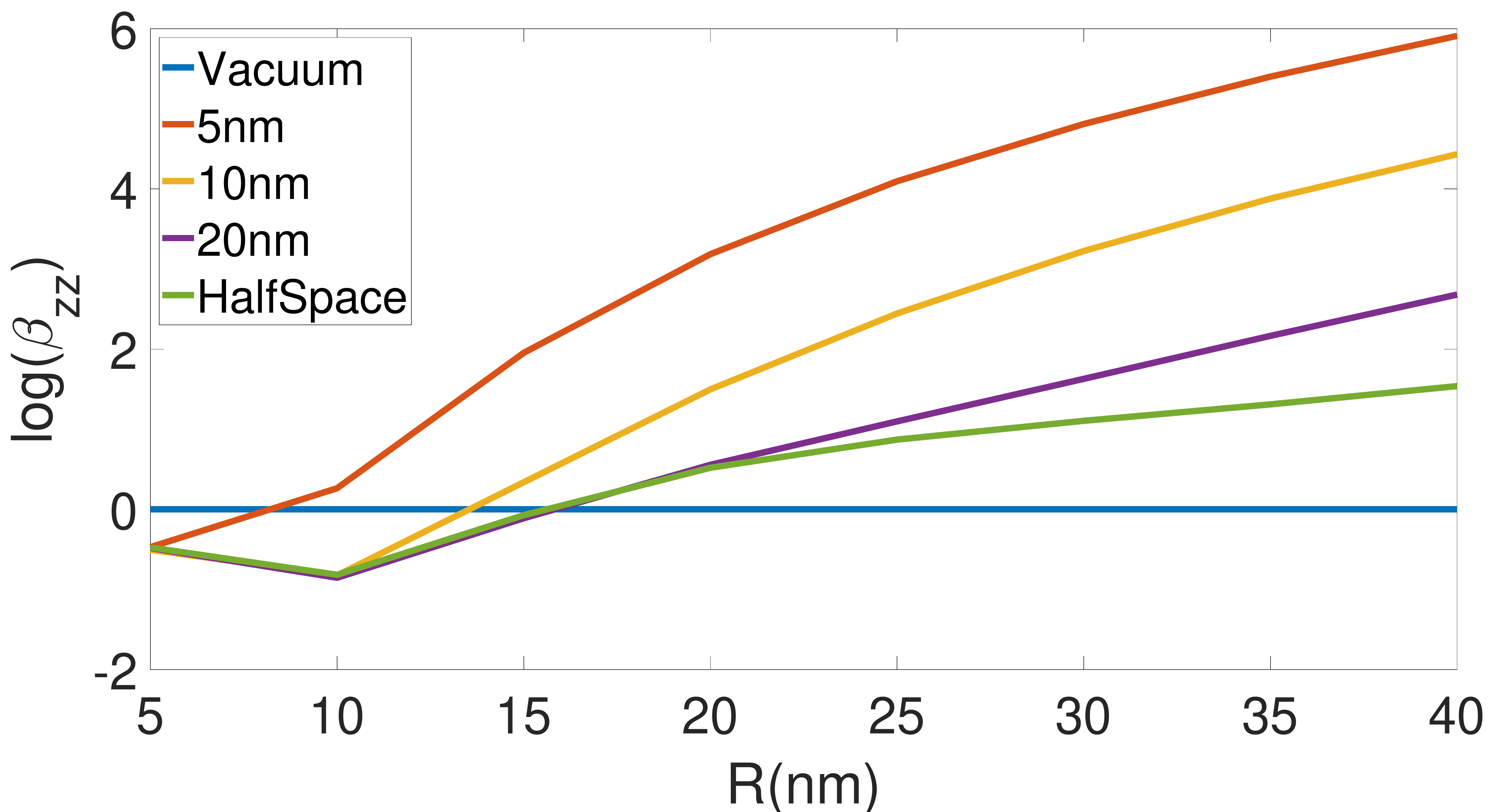}}\label{Beta_z}
    \caption{The change of $\beta$ and $\gamma_{MC}/\gamma_{0}$ with $R$ in the $XX$, $YY$ and $ZZ$ configurations.}\label{ration_gamma}
\end{figure}

The energy transfer is strongly coupled with the evanescent EM waves in the vicinity of the thin metallic film. The evanescent EM waves have two components: the radiation by the dipole of chromophore molecule, and the scattering of radiation by the interface of the metallic thin films. The two components of EM waves are described by the vacuum DGF $\hat{\mathbf{G}}^{0}$ and the scattering DGF $\hat{\mathbf{G}}^{sc}$ respectively. (The scattering DGF of the metallic thin film is presented in Appendix~\ref{ADGF}). The surface of metallic thin film breaks the spatial rotational symmetry in the vacuum space, resulting in different coupling strengths between chromophore molecules with different polarization directions, and in turn affects the rate of MC-FRET. The two-to-two MC structure is used to study the influence of the surface scattering on the MC systems with different molecular polarization directions. The configuration of the two-to-two MC structure is shown in Fig.~\ref{2to2sketch}. In the two-to-two structure, the distance between the two acceptors is 5 nm, as well as for the distance between the two donors.  The two-to-two MC structure has one geometric factor, R, that is the distance between the centers of the donors and acceptors.
Since $|\hat{\mathbf{G}}^{0}_{ii}|\gg|\hat{\mathbf{G}}^{sc}_{ii}|$ when the distance R is small enough, the scattering wave component has no influence on MC-FRET. As R increases, the scattering components gradually inhibits MC-FRET. Since $\hat{\mathbf{G}}^{0}_{ii}$ decays more rapidly than $\hat{\mathbf{G}}^{sc}_{ii}$, beyond a certain distance, the interface of the metallic thin film has gaining effect on MC-FRET.
In the two-to-two MC structure, the polarization orientations have the XX, YY and ZZ three configurations respectively. For each configuration, the ratio $\frac{\gamma_{MC}}{\gamma_0}$ shows how the interface of metallic thin film affects the MC-FRET rate in benchmark to the MC-FRET rate in vacuum. 
The ratio $\frac{\gamma_{MC}}{\gamma_0}$ clearly deceases in the small R then increase for the $XX$ and $YY$ configurations. However, for the $ZZ$ configuration, the ratio $\frac{\gamma_{MC}}{\gamma_0}$ basically increase monotonically. Fig.~\ref{ration_gamma} shows that the ratio $\frac{\gamma_{MC}}{\gamma_0}$ is smaller than 1 when R is small and further increases and becomes larger than 1 when R increases.

In the two-to-two MC structure, the two donor chromophores and two acceptor chromophores are aligned in the X axis. Therefore, the two-to-two MC structure has $SO2$ symmetry along the X direction in the vacuum. The dyadic Green function in stay unchanged under the rotation operator around the X direction in the $YZ$ plane.  Therefore, there are two essential components, one is parallel to X axis and the other perpendicular to the X axis in the $YZ$ plane. 
According to the Dyadic Green function in Eq~(\ref{G0}), the can be re-written to be 
\begin{equation}\label{G0}
\hat{\mathbf{G}}^{0}=\left[\begin{array}{ccc}
     g_{\parallel}& 0& 0\\
     0&  g_{\perp}& 0\\
      0& 0&g_{\perp}
\end{array}
\right]
\end{equation}
where, the parallel component corresponding to the $XX$ component is 
\begin{equation}\label{gpar}
g_{\parallel}=\frac{\exp(i\mathit{k}\mathit{R})}{4\pi R} \left[ \left(1+\frac{i\mathit{k}\mathit{R}-1}{\mathit{k}^2\mathit{R}^2}  \right) 
+ \frac{3-3i \mathit{k} \mathit{R}-\mathit{k}^2\mathit{R}^2}{\mathit{k}^2\mathit{R}^2} 
\right] ,
\end{equation}
and the perpendicular component corresponding to the $YY$ and $ZZ$ components  is
\begin{equation}\label{gperp}
g_{\perp}=\frac{\exp(i\mathit{k}\mathit{R})}{4\pi R} \left(1+\frac{i\mathit{k}\mathit{R}-1}{\mathit{k}^2\mathit{R}^2}  \right) ,
\end{equation}
With the $SO2$ symmetry operator around the X axis defined as,
\begin{equation}\label{rotate_operator}
\hat{\mathbf{R}}_{X}(\theta)=\left[\begin{array}{ccc}
     1& 0& 0\\
     0&  cos(\theta)& sin(\theta)\\
      0& -sin(\theta)&cos(\theta)
\end{array}
\right]
\end{equation}
Therefore the $YY$  and $ZZ$ configurations are equivalent in the vacuum since   
\begin{equation}\label{rotate_operator}
\hat{\mathbf{R}}_{X}^{T}(\theta)\hat{\mathbf{G}}^{0}
\hat{\mathbf{R}}_{X}(\theta)=\hat{\mathbf{G}}^{0}
\end{equation}
However, when the interface of metallic thin film exists, the $SO2$ symmetry breaks down. The MC-FRET rates have different R distance dependence for the $XX$, $YY$, and $ZZ$ configurations.

It can be seen from the spectral overlapping in Fig.~\ref{spectrum} is mainly around 420 nm, so the scattering DGF at 420 nm plays an important role in MC-FRET. To quantitatively describe the influence of the scattering wave on MC-FRET at a specific frequency $\omega$, the ratio $\beta_{ii}$ is defined as,
\begin{equation}\label{beta}
    \beta_{ii}(\mathbf{r}_A;\mathbf{r}_D;\omega)=\frac{|\hat{\mathbf{G}}^{sc}_{ii}(\mathbf{r}_A;\mathbf{r}_D;\omega)+\hat{\mathbf{G}}^{0}_{ii}(\mathbf{r}_A;\mathbf{r}_D;\omega)|^2}{|\hat{\mathbf{G}}^{0}_{ii}(\mathbf{r}_A;\mathbf{r}_D;\omega)|^2}\ (i=x,y,z)
\end{equation}
When $\beta_{ii}>1$ it means that the interface enhance MC-FRET at this frequency in $ii$ configuration, and vice versa, it suppresses this process. For $i=x,y,z$, the $\beta_{ii}$ are shown in Fig~\ref{ration_gamma}. By comparison, it is clear that $\beta$ and $\gamma_{MC}/\gamma_{0}$ have similar patterns that change with $R$. 

\begin{figure}[H]
    \centering
    \subfigure[ $I_{\pm}(\kappa)$ vs $T_{-}^{p}(\kappa)$]{\includegraphics[width=0.496\textwidth]{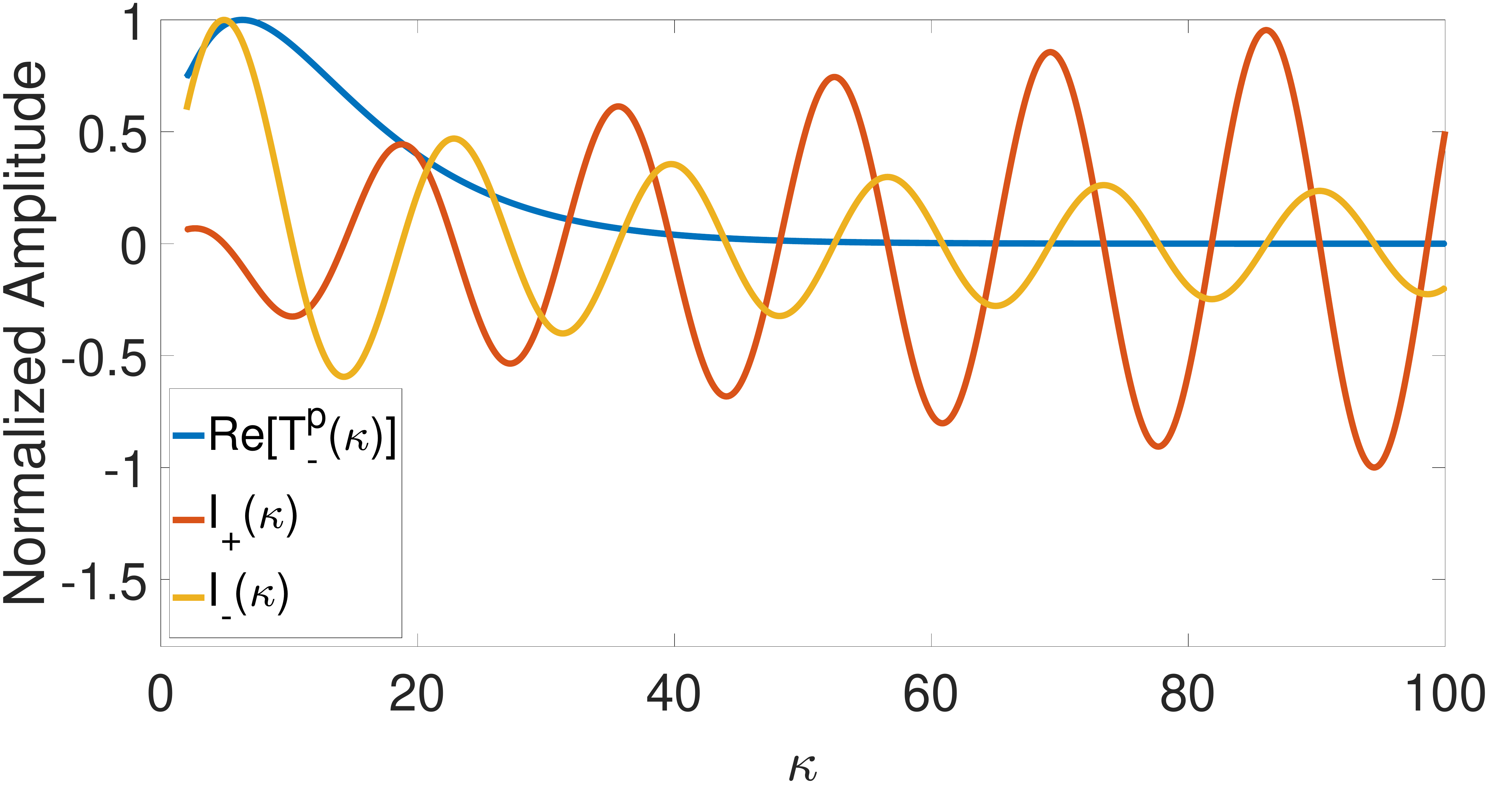}}\label{sign_xy}
    \subfigure[$I_{0}(\kappa)$ vs $T_{+}^{p}(\kappa)$ ]{\includegraphics[width=0.496\textwidth]{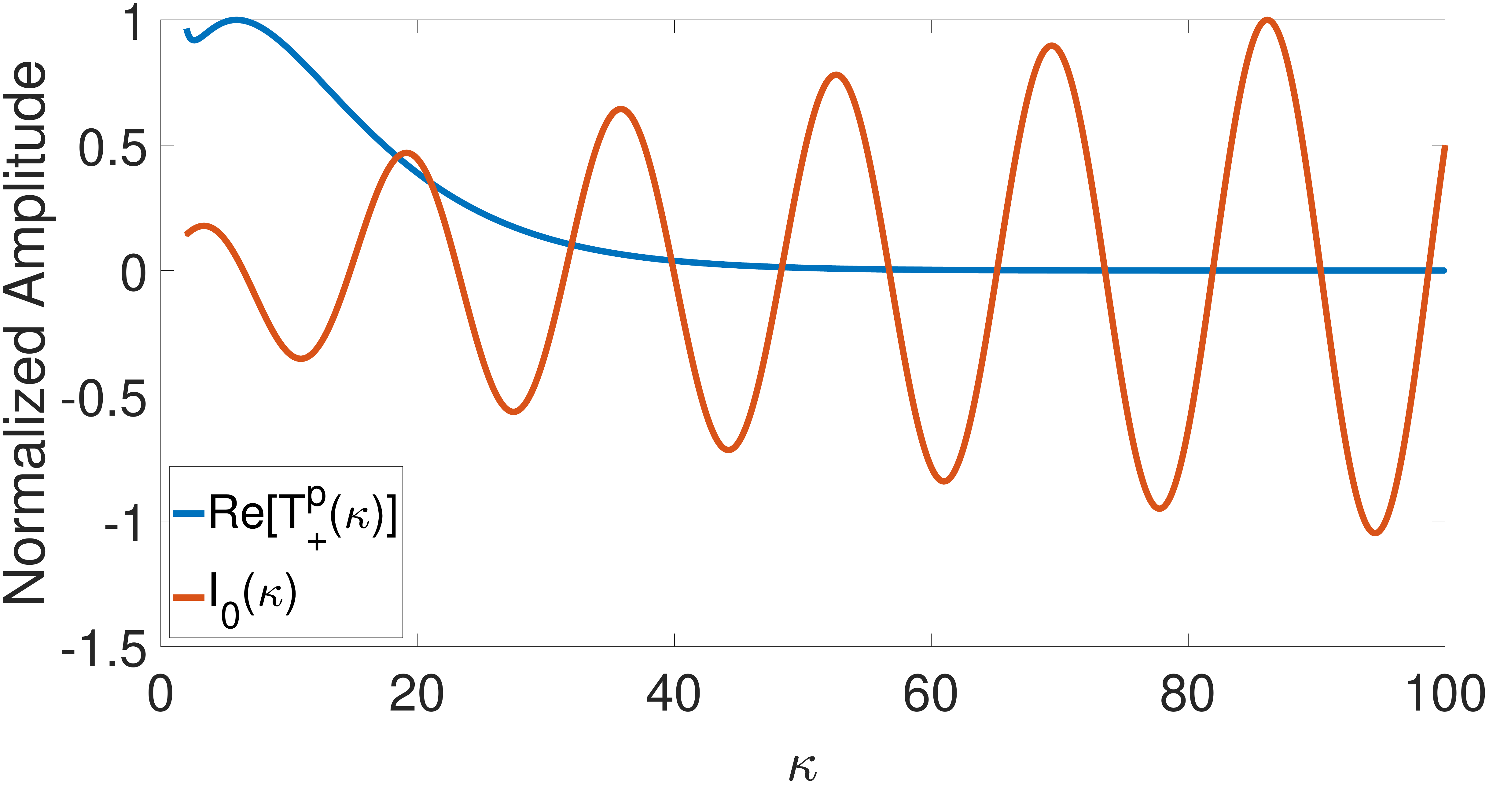}}\label{sign_zz}%
    \caption{The overlapping between the $I_{\pm}(\kappa)$ and the $Re[T_{-}^{p}(\kappa)]$ for the interface of silver halfspace  (a), and the overlapping between $I_{0}(\kappa)$ and the $Re[T_{+}^{p}(\kappa)]$ of the interface of the silver halfspace (b) for R=25 nm, z=5 nm, $\lambda$=420 nm.}\label{SignOfG}
\end{figure}

 In the two-to-two MC system with R = 25 nm,  the contribution of the evanescent wave with large horizontal momentum is dominant in the diagonal components of the scattering DGF $\hat{\mathbf{G}}^{sc}_{xx}$, $\hat{\mathbf{G}}^{sc}_{yy}$ and $\hat{\mathbf{G}}^{sc}_{zz}$ at $\lambda =$420 nm ($\kappa\gg1$, see Appendix~\ref{ADGF}.). The influence of S wave and propagating wave can be ignored to obtain reduced scattering DGF (refer to Eq.~\ref{DGFzz1} to \ref{I1} in Appendix~\ref{red}). The reduced scattering DGF can be artificially decomposed into the integral of the product of two factors, the scattering factor  $T_{\pm}^{p}(\kappa)$ and interference factor $I_{0,\pm}(\kappa)$. The scattering factor $T_{\pm}^{p}(\kappa)$ describes the projected scattering EM field of a single ${\kappa}$-mode and interference factor $I_{0,\pm}(\kappa)$ describes the interference of all ${\kappa}$-modes. The overlapping between these two factors determines the total scattering EM field at $\lambda =$420 nm. The changes of the $I_{0,\pm} (\kappa)$ and the $T_{\pm}^{p} (\kappa)$ with the $\kappa$ momentum are shown in Fig.~\ref{SignOfG}.

\section{Concluding Remarks}\label{cr}
The classic  DGF approach to MC-FRET provides a microscopic understanding of the interaction between the MC systems and the evanescent electromagnetic field. The MC-FRET rates under electromagnetic fluctuations show complicated distance dependence behavior. Particularly, the MC-FRET rate of ring structures shows complicated distance dependence in the vicinity of the thin silver film. In summary, we have presented a generalized MC-FRET classic approach to study EET n the multichromophoric systems in the vicinity of the metallic thin film. The classic DGF approach to EET has provided new evidence for efficient and dispersive energy transfer dynamics caused by the interaction between MC systems and the evanescent EV waves above metallic thin films. The simulation results suggest complex MC-FRET rate distance-dependence behavior. In addition, polarization orientations also play a critical role in the MC-FRET rate modulation. The arrangement of chromophores and the interaction between MC structures and evanescent EM modes in the vicinity of the metallic thin film can control the MC-FRET rate in the MC structures. 

We conclude this paper by discussing the implications of our results. The generalized MC-FRET approach can be used in nanophotonic crystals. The nanophotonic metasurface and materials can provide more freedom to control the evanescent EV waves and modulate MC-FRET. In the future, we can study how to design the nanophotonic metasurface structure to enhance the resonant energy transfer. Understanding the relationship between the MC structure and its excitation energy transfer properties is of fundamental importance in many applications, including the development of next-generation photovoltaics. Computational insights into energy transport dynamics can be gained by leveraging the DGF approach numerically. Our future work is to design the AI inverse design framework to identify optimal nanophotonic metasurface and materials to enhance the energy transfer efficiency of organic solar cells and the quantum efficiency of organic light-emitting diode (LED).

\section*{Acknowledgment}
Xin Chen acknowledges the funding support from the National Natural Science Foundation of China under grant No. 21773182 and the support of HPC Platform, Xi’an Jiaotong University.

\appendix

\section{Scattering DGF of Hyperbolic Multi-layer Thin Films}\label{ADGF}
The Scattering DGF of hyperbolic multi-layer thin film are widely studied and can be expressed analytically. A dipole $\mathbf{P}(\mathbf{r}_0)$ with coordinates $\mathbf{r}_0=(0,0,z_0)$ above the interface,can be expanded into a series of sheets of polarization by Fourier expansion.
\begin{equation}\label{Fourier_expansion}
    \mathbf{p}\delta(\mathbf{r}-\mathbf{r}_0)=\frac{\mathbf{p}}{(2\pi)^2}\int_{-\infty}^{\infty}\int_{-\infty}^{\infty}e^{i(\kappa_xx+\kappa_yy)}{d\kappa_x}{d\kappa_y}
\end{equation}
The spatial distribution of dipoles in each polarization sheet is 
\begin{equation}\label{distribution}
\mathbf{p}_{\kappa_x,\kappa_y}(\mathbf{r})=\mathbf{p}\delta(z-z_0)e^{i(\kappa_xx+\kappa_yy)}
\end{equation}
In cylindrical coordinates, each polarization sheet can be relabeled by $\kappa$ and $\varphi$
\begin{equation}\label{distribution}
\mathbf{p}_{\kappa,\varphi}(\mathbf{r})=\mathbf{p}\delta(z-z_0)e^{i(\kappa{cos\varphi}x+\kappa{sin\varphi}y)}
\end{equation}
Where
\begin{equation}\label{kappa}
\kappa=\sqrt{\kappa_x^2+\kappa_y^2}
\end{equation}
and
\begin{equation}\label{phi}
\varphi=arctan(\kappa_y/\kappa_x).
\end{equation}
The wave vector of the electromagnetic wave excited by each polarization sheet is $\mathbf{\nu}=({\kappa}cos{\varphi},{\kappa}sin{\varphi},p)(\frac{\omega}{c})$, where $p=\sqrt{1-{\kappa}^2}$ gives modes of propagating waves if $\kappa<1$, and $p=i\sqrt{{\kappa}^2-1}$ gives the modes of evanescent waves if $\kappa>1$. where $\omega$ is frequency, c is the speed of light. In addition to directly exciting 
EM waves into the vacuum, the polarization sheet will also be scattered by the interface. Considering that there are two kinds of polarization, s and p, for each polarization sheet, the electric field at $\mathbf{r}_A$ can be directly calculated with Maxwell's equations\cite{sipe1987new}:
\begin{equation}\label{phi}
\mathbf{E}(\kappa,\varphi,\mathbf{r}_A)=\frac{i{\omega}^3{\mu_0}}{4\pi{p}{c}}(\hat{s}\hat{s}+\hat{p}_{+}\hat{p}_{+})e^{i\mathbf{\nu}\cdot\mathbf{r}_M}\cdot\mathbf{p}(\mathbf{r}_0)+\frac{i{\omega}^3{\mu_0}}{4\pi{p}{c}}(\hat{s}r^{s}\hat{s}+\hat{p}_{+}r^{p}\hat{p}_{-})e^{i\mathbf{\nu}\cdot\mathbf{r}_M}\cdot\mathbf{p}(\mathbf{r}_0)
\end{equation}
The first term and the second term correspond to the vacuum Green's function and the scattering Green's function of the polarization sheet, respectively. Where
$\mathbf{r}_M = (\mathbf{r}_{A, x},\mathbf{r}_{A, y}, \mathbf{r}_{A,z}+z_0)$ is the distance from the acceptor to the mirror image of the donor. $r^{s}$ and $r^{p}$ are the Fresnel reflection factors for the interface of the medium where the upper indices of s and p indicates the two polarization. The direction of the s and p polarization are:
\begin{equation}\label{s}
    \hat{s}\equiv(sin(\varphi),-cos(\varphi),0),
\end{equation}
and
\begin{equation}\label{pin}
    \hat{p}_{\pm}\equiv({\pm}cos(\varphi)p,{\pm}sin(\varphi)p,\kappa),
\end{equation}
According to Eq.~\ref{Fourier_expansion}, the field excited by a point dipole is equal to the integral of the polarization sheet of all modes. The electric field in the scattering part can be written as
\begin{equation}\label{Esc}
\mathbf{E}^{sc}(\mathbf{r}_A)=\frac{i{\omega}^3{\mu_0}}{8{\pi}^2c}PV\int_{0}^{\infty}\int_{0}^{2\pi}\frac{\kappa}{p}(\hat{s}r^{s}\hat{s}+\hat{p}_{+}r^{p}\hat{p}_{-})e^{i\mathbf{\nu}\cdot\mathbf{r}_M}\cdot\mathbf{p}(\mathbf{r}_0)d{\varphi}d{\kappa},
\end{equation}
Considering the definition of Green's function and the spatial translation symmetry in the XY plane, the scattering DGF above the infinite halfspace can be obtained,
\begin{equation}\label{DGFsingle}
    \hat{\mathbf{G}}^{sc}(\mathbf{r}_A;\mathbf{r}_D;\omega)=\frac{i}{8\pi^2}(\frac{\omega}{c})PV\int_{0}^{\infty}\int_{0}^{2\pi}\frac{\kappa}{p}(\hat{s}r^{s}\hat{s}+\hat{p}_{+}r^{p}\hat{p}_{-})e^{i\mathbf{\nu}\cdot\mathbf{r}_M}d{\varphi}d{\kappa},
\end{equation}
At this time $\mathbf{r}_M = (\mathbf{r}_{A, x}-\mathbf{r}_{D,x},\mathbf{r}_{A, y}- \mathbf{r}_{D,y}, \mathbf{r}_{A,z}+\mathbf{r}_{D, z})$. For simplicity, we assume that the direction of $\mathbf{R}$ vector from the donor and the acceptor is the X direction. Considering the spatial rotation in-variance of the multi-layer system in the Z direction, there are only four nonzero components in $3\times3$ dydaic green's function. Using the Jacobi–Anger expansion, these four nonzero components can be obtained as,
\begin{equation}\label{DGFxx}
    \begin{split}
        \hat{\mathbf{G}}^{sc}_{xx}(\mathbf{r}_A;\mathbf{r}_D;w) =& \frac{i}{8{\pi}}(\frac{w}{c})PV\int_{0}^{\infty}\frac{\kappa}{p}r^{s}(\kappa)e^{2ip{\omega}z/c}\left[J_0(\frac{\kappa\omega{R}}{c})+J_2(\frac{\kappa\omega{R}}{c})\right]d{\kappa}\\
        &-\frac{i}{8{\pi}}(\frac{w}{c})PV\int_{0}^{\infty}{\kappa}{p}r^{p}(\kappa)e^{2ip{\omega}z/c}\left[J_0(\frac{\kappa\omega{R}}{c})-J_2(\frac{\kappa\omega{R}}{c})\right]d{\kappa}
    \end{split}
\end{equation}
\begin{equation}\label{DGFyy}
    \begin{split}
        \hat{\mathbf{G}}^{sc}_{yy}(\mathbf{r}_A;\mathbf{r}_D;w) =& \frac{i}{8{\pi}}(\frac{w}{c})PV\int_{0}^{\infty}\frac{\kappa}{p}r^{s}(\kappa)e^{2ip{\omega}z/c}\left[J_0(\frac{\kappa\omega{R}}{c})-J_2(\frac{\kappa\omega{R}}{c})\right]d{\kappa}\\
        &-\frac{i}{8{\pi}}(\frac{w}{c})PV\int_{0}^{\infty}{\kappa}{p}r^{p}(\kappa)e^{2ip{\omega}z/c}\left[J_0(\frac{\kappa\omega{R}}{c})+J_2(\frac{\kappa\omega{R}}{c})\right]d{\kappa}
    \end{split}
\end{equation}
\begin{equation}\label{DGFzz}
    \hat{\mathbf{G}}^{sc}_{zz}(\mathbf{r}_A;\mathbf{r}_D;\omega)=\frac{i}{4{\pi}}(\frac{\omega}{c})PV\int_{0}^{\infty}\frac{\kappa^3}{p}r^{p}(\kappa)J_0(\frac{\kappa\omega R}{c})e^{2ip{\omega}z/c}d{\kappa}
\end{equation}
\begin{equation}\label{DGFxz}
    \hat{\mathbf{G}}^{sc}_{xz}(\mathbf{r}_A;\mathbf{r}_D;\omega)=\frac{1}{4{\pi}}(\frac{\omega}{c})PV\int_{0}^{\infty}{\kappa^2}r^{p}(\kappa)J_1(\frac{\kappa\omega R}{c})e^{2ip{\omega}z/c}d{\kappa}=-\hat{\mathbf{G}}^{sc}_{zx}
\end{equation}
where $R \equiv \sqrt{(\mathbf{r}_{A, x}-\mathbf{r}_{D, x})^2+(\mathbf{r}_{A, y}-\mathbf{r}_{D, y})^2}$ is the horizontal distance betweem the donor and the acceptor.
And $z\equiv\frac{1}{2}(r_{a}^{z}+r_{d}^{z})$ is the average distance from the donor and acceptor to the surface of the multilayer structure. $J_n(\kappa)$ is the Bessel functions of n order.
For single-layer film system, we need to substitute $R^{s/p}$ for $r^{s/p}$. Where,
\begin{equation}\label{RofTheFilm}
    R^{s/p}=r^{s/p}\frac{1-e^{2ik_{mz}d}}{1-(r^{s/p})^2e^{2ik_{mz}d}}
\end{equation}
The $k_{mz}$ in Eq.\ref{RofTheFilm} is the z component of the wave vector in the film, and the $d$ is the thickness of the film. 

\section{Reduced Scattering DGF}\label{red}
The diagonal $zz$ component in $\hat{\mathbf{G}}$ a has no the contribution from the $s$ wave therefore can ignore the contribution from the propagating waves. The reduced $zz$ component of the scattering DGF in Eq.~\ref{DGFzz} can be expressed by ignoring the contribution of $s$ wave,
\begin{equation}\label{DGFzz1}
    \hat{\mathbf{G}}^{sc}_{zz}(\mathbf{r}_A;\mathbf{r}_D;\omega) \approx\frac{1}{8{\pi^2}}(\frac{w}{c})PV\int_{1}^{\infty}T_{+}^{p}(\kappa)I_0(\frac{\kappa\omega R}{c})d{\kappa}
\end{equation}
where
\begin{equation}\label{Sm}
    T_{+}^{p}(\kappa)={\kappa}(\frac{\kappa}{|p|})^{+1}r^{p}(\kappa)e^{-2|p|{\omega}z/c},
\end{equation}
and
\begin{equation}\label{I0}
    I_0(\kappa)=2\pi{\kappa}{J_0(\frac{\kappa\omega{R}}{c})}
\end{equation}
In the $zz$ component of the scattering DGF in Eq.~\ref{DGFzz1}, there are two factor, the scattering factor  $T_{+}^{p}(\kappa)$ and interference factor $I_0(\kappa)$. The scattering factor $T_{+}^{p}(\kappa)$ describes the reflected EM field of a single ${\kappa}$-mode) and interference factor $I_0(\kappa)$ describes the interference of all ${\kappa}$-modes. The overlapping between $I_0$ and  $T_{+}^{p}(\kappa)$ factors determines how the evanescent near field can enhance the FRET.
However, the remaining three reduced components in the scattering DGF share the similar simplified expression
by ignoring the contribution of $s$ wave.
When the distance between molecules is much smaller than the wavelength, the component with larger horizontal momentum($\kappa\gg1$) dominates, and at this time $\kappa|p|\gg\frac{\kappa}{|p|}$, we can further ignore the contribution of the s-wave as in Eq.~\ref{DGFxx} and Eq.~\ref{DGFyy} (refer to Appendix~\ref{ADGF}). Thus, the four reduced $xx$, $yy$, $xz$ and $zx$ components can be approximated in terms of  $T_{\pm}^p(\kappa)$, $I_{\pm} (\kappa)$ and $I_{1}(\kappa)$ as,
\begin{equation}\label{DGFxx1}
    \hat{\mathbf{G}}^{sc}_{xx}(\mathbf{r}_A;\mathbf{r}_D;\omega) \approx \frac{1}{8{\pi}^2}(\frac{\omega}{c})PV\int_{1}^{\infty}T_{-}^{p}(\kappa)I_{-}(\kappa)d{\kappa},
\end{equation}
\begin{equation}\label{DGFyy1}
    \hat{\mathbf{G}}^{sc}_{yy}(\mathbf{r}_A;\mathbf{r}_D; \omega) \approx \frac{1}{8{\pi}^2}(\frac{\omega}{c})PV\int_{1}^{\infty}T_{-}^{p}(\kappa)I_{+}(\kappa)d{\kappa},
\end{equation}
\begin{equation}\label{DGFxz}
    \hat{\mathbf{G}}^{sc}_{xz}(\mathbf{r}_A;\mathbf{r}_D; \omega) \approx \frac{i}{8{\pi}^2}(\frac{\omega}{c})PV\int_{1}^{\infty}T_{-}^{p}(\kappa)I_{1}(\kappa)d{\kappa}=-\hat{\mathbf{G}}^{sc}_{zx}(\mathbf{r}_A;\mathbf{r}_D; \omega),
\end{equation}
where
\begin{equation}\label{Sp}
    T_{\pm}^{p}(\kappa)={\kappa}(\frac{\kappa}{|p|})^{\pm1}r^{p}(\kappa)e^{-2|p|{\omega}z/c},
\end{equation}
\begin{equation}\label{Ip}
    I_{+}(\kappa)=\int_{0}^{2\pi}{\kappa}sin^2{\varphi}e^{i\kappa\omega{cos(\varphi)}R/c}d{\varphi}=\pi{\kappa}[J_0(\frac{\kappa\omega{R}}{c})+J_2(\frac{\kappa\omega{R}}{c})],
\end{equation}
\begin{equation}\label{Im}
    I_{-}(\kappa)=\int_{0}^{2\pi}{\kappa}cos^2{\varphi}e^{i\kappa\omega{cos(\varphi)}R/c}d{\varphi}=\pi{\kappa}[J_0(\frac{\kappa\omega{R}}{c})-J_2(\frac{\kappa\omega{R}}{c})],
\end{equation}
\begin{equation}\label{I1}
    I_{1}(\kappa)=\int_{0}^{2\pi}{\kappa}cos{\varphi}e^{i\kappa\omega{cos(\varphi)}R/c}d{\varphi}=i2{\pi}{\kappa}J_1(\frac{\kappa\omega{R}}{c}).
\end{equation}

\section{Generalized MC-FRET Simplification}\label{GMC-FRET-Simp}
The first part of the generalized MC-FRET rate in Eq.~\ref{Qvector} is reformatted and simplified with the conjugate symmetry properties of DGF in the region of $[0, \infty]$ to be,
\begin{eqnarray}
    &&\int_{-\infty}^{\infty}\; d\omega\; \omega^3\;  \hat{\mathbf{G}}^{\mathrm{A}*} (\omega) \cdot \mathbf{P}_{\mathrm{A}}^*(\omega) \cdot \mathbf{P}_{\mathrm{A}}(\omega)
      \\ \nonumber
    &=&  \int_{0}^{\infty}\; d\omega\;\omega^3 \;  \hat{\mathbf{G}}^{\mathrm{A}*} (\omega) \cdot \mathbf{P}_{\mathrm{A}}^*(\omega) \cdot \mathbf{P}_{\mathrm{A}}(\omega) + \int_{-\infty}^{0}\;d\omega \;\omega^3\; \hat{\mathbf{G}}^{\mathrm{A}*} (\omega) \cdot \mathbf{P}_{\mathrm{A}}^*(\omega) \cdot \mathbf{P}_{\mathrm{A}}(\omega)  \\ \nonumber
    &=&
    \int_{0}^{\infty} \;d\omega\; \omega^3 \;  \hat{\mathbf{G}}^{\mathrm{A}*} (\omega) \cdot \mathbf{P}_{\mathrm{A}}^*(\omega) \cdot \mathbf{P}_{\mathrm{A}}(\omega) - \int^{\infty}_{0} \;d\omega\; \omega^3 \;  \hat{\mathbf{G}}^{\mathrm{A}*} (-\omega) \cdot \mathbf{P}_{\mathrm{A}}^*(-\omega) \cdot \mathbf{P}_{\mathrm{A}}(-\omega)  \\ \nonumber
    &=&
    \int_{0}^{\infty}\; d\omega\; \omega^3 \;  \hat{\mathbf{G}}^{\mathrm{A}*} (\omega) \cdot \mathbf{P}_{\mathrm{A}}^*(\omega) \cdot \mathbf{P}_{\mathrm{A}}(\omega) - \int^{\infty}_{0}\; d\omega\; \omega^3 \;\hat{\mathbf{G}}^{\mathrm{A}} (\omega) \cdot \mathbf{P}_{\mathrm{A}}(\omega) \cdot \mathbf{P}_{\mathrm{A}}^*(\omega) \\ \nonumber
    &=&
    2 i \int_{0}^{\infty}  \; d\omega  \;  \omega^3  \; \textbf{Im} \Big(  \hat{\mathbf{G}}^{\mathrm{A}} (\omega) \cdot \mathbf{P}_{\mathrm{A}}^*(\omega) \cdot \mathbf{P}_{\mathrm{A}}(\omega) \Big)
\end{eqnarray}
Similarly, the second part in Eq.~\ref{Qvector} becomes,
\begin{eqnarray}
    &&\int_{-\infty}^{\infty} \; d\omega \; \omega^3 \;\hat{\mathbf{G}}^{\mathrm{AD}*} (\omega) \cdot \mathbf{P}_{\mathrm{D}}^*(\omega) \cdot \mathbf{K}(\omega)^{-1} \cdot \hat{\mathbf{G}}^{\mathrm{AD}}(\omega) \cdot \mathbf{P}_{\mathrm{D}}(\omega) \\ \nonumber
    &=& 2 i \int_{0}^{\infty} d\omega \; \omega^3  \; \textbf{Im}  \Big( \hat{\mathbf{G}}^{\mathrm{AD}*} (\omega) \cdot \mathbf{P}_{\mathrm{D}}^*(\omega) \cdot \mathbf{K}(\omega)^{-1} \cdot \hat{\mathbf{G}}^{\mathrm{AD}}(\omega)  \mathbf{P}_{\mathrm{D}}(\omega) \Big),
\end{eqnarray}
due to the conjugate symmetry $\mathbf{K}(-\omega)^{-1} = \mathbf{K}^*(\omega)^{-1}$.
In comparison to the conventional FRET expression, the MC-FRET rate is defined as, 
\begin{eqnarray}\label{Qvectorfinal1}
    \gamma_{MC}
    &= & 2 \mu_0 \int_{0}^{\infty} d\omega \;  \omega^3  \; \textbf{Im}  \Big(  \hat{\mathbf{G}}^{\mathrm{A}*} (\omega) \cdot \mathbf{P}_{\mathrm{A}}^*(\omega) \cdot \mathbf{P}_{\mathrm{A}}(\omega) \\ \nonumber
    & + & \hat{\mathbf{G}}^{\mathrm{AD}*} (\omega) \cdot \mathbf{P}_{\mathrm{D}}^*(\omega)\cdot \mathbf{K}(\omega)^{-1} \cdot \hat{\mathbf{G}}^{\mathrm{AD}}(\omega) \cdot \mathbf{P}_{\mathrm{D}}(\omega)   \Big)
\end{eqnarray}

\section{Three-Fold Ring MC-FRET}\label{3fold}
The donor and acceptor induced polarizability vectors can be expressed as $\mathbf{P}_{\mathrm{A}}(\omega)=P_A(\omega)\mathbf{v}_A$ and $\mathbf{P}_{\mathrm{D}}(\omega)={P}_D(\omega)\mathbf{v}_D$ where $P_A(\omega)$ is the induced polarizability of the acceptor chromophore, $P_D(\omega)$ the induced polarizability of the donor chromophore, $\mathbf{v}_A=\mathbf{v}_1^A\oplus \cdots \oplus \mathbf{v}_i^A \oplus \cdots \oplus \mathbf{v}_n^A$ the direct sum of the unit vector $\mathbf{v}_i^A$ for each chromophore molecule in the acceptor ring aggregator, and $\mathbf{v}_D=\mathbf{v}_1^D\oplus \cdots \oplus \mathbf{v}_i^D \oplus \cdots \oplus \mathbf{v}_n^D$ the direct sum of the unit vector $\mathbf{v}_i^D$ for each chromophore molecule in the donor ring aggregator. At the end, the MC-FRET rate can be reduced to the following form in wavelength $\lambda$,
\begin{eqnarray}\label{Qvectorring}
    \gamma_{MC}
    &= & 2 \int_{0}^{\infty} d\lambda \frac{(2\pi{c})^4}{\lambda^5} P_D(\lambda)^2  \\ \nonumber
    & &  \textbf{Im} \left\{ \mathbf{K}^*(\lambda)^{-1} \hat{\mathbf{G}}^{\mathrm{AD}*}(\lambda) \mathbf{v}_D \cdot \hat{\mathbf{G}}^{\mathrm{A}}(\lambda)  \mathbf{K}(\lambda)^{-1}  \hat{\mathbf{G}}^{\mathrm{AD}}(\lambda)  \mathbf{v}_D
    +
    \mathbf{K}^*(\lambda)^{-1}\hat{\mathbf{G}}^{\mathrm{AD}*}(\lambda) \mathbf{v}_D  \cdot    \hat{\mathbf{G}}^{\mathrm{AD}} (\lambda)  \mathbf{v}_D  \right\},
\end{eqnarray}
where $P_D(\lambda)^2$ is the emission spectrum of the donors.

\bibliography{this}   
\end{document}